\documentclass[11pt]{article}

\usepackage[dvips,
colorlinks=true,
linkcolor=black,
citecolor=blue,
urlcolor=black,
bookmarks=true,
bookmarksnumbered=true,
bookmarkstype=toc]{hyperref}
\usepackage[a4paper,centering,text={6in,9in}]{geometry}
\usepackage{amsmath,amssymb,amsbsy}
\usepackage{mathrsfs}
\usepackage{bm}
\usepackage{mathptmx}
\usepackage{subfigure}
\usepackage{epic,eepic,xcolor}
\usepackage{cite}
\makeatletter

\@addtoreset{equation}{section}
\makeatother
\DeclareMathAlphabet{\mathcal}{OMS}{cmsy}{m}{n}

\title{
\vskip -3em
\begin{flushright}
\small
HRI-P-12-10-001
\end{flushright}
\vskip 5em
\textbf{\large Parasupersymmetry in Quantum Graphs}
}
\author{
Satoshi Ohya
\\[1em]
\textit{\small Harish-Chandra Research Institute}\\
\textit{\small Chhatnag Road, Jhusi, Allahabad 211 019, India}\\[1ex]
\texttt{\small E-mail:\href{mailto:ohya@hri.res.in}{ohya@hri.res.in}}
}
\date{\small (Dated: \today)}

\begin{document}
\maketitle
\thispagestyle{empty}
\begin{abstract}
We study hidden parasupersymmetry structures in purely bosonic quantum mechanics on compact equilateral graphs.
We consider a single free spinless particle on the graphs and show that the Huang-Su parasupersymmetry algebra is hidden behind degenerate spectra.
\end{abstract}
\vskip 2em
\tableofcontents

\newpage
\section{Introduction} \label{sec:1}
Parasupersymmetry is a generalization of supersymmetry to parastatistics: It is a symmetry between bosons and parafermions.
Historically, parasupersymmetry was first introduced in non-relativistic quantum mechanics by Rubakov and Spiridonov \cite{Rubakov:1988mv}, and nowadays there are two different formulations of parasupersymmetric quantum mechanics.
One is parasupersymmetric quantum mechanics of order $2$ proposed by Rubakov and Spiridonov \cite{Rubakov:1988mv}, which is extended to arbitrary order by Khare \cite{Khare:1992gy}, Tomiya \cite{Tomiya:1992} and Huang and Su \cite{Huang:2010zzc},\footnote{Higher order parasupersymmetry in terms of a single hermitian parasupercharge was also investigated in \cite{Durand:1991ec}.} and the other is that proposed by Beckers and Debergh \cite{Beckers:1990av}, which is extended to arbitrary order by Chenaghlou and Fakhri \cite{Chenaghlou:2002zm}.
Common ingredients of these two parasupersymmetric quantum mechanics are Hamiltonian operator $H$, parafermion number operator $N_{\text{PF}}$, nilpotent parafermionic charge (parasupercharge) $Q^{+}$ and its adjoint $Q^{-}$ that satisfy $(Q^{\pm})^{p+1} = 0$, where $p \in \mathbb{N}$ is the order of parafermion.
Parasupersymmetric quantum mechanics is characterized by the parasupersymmetry algebra that consists of the commutation relations
\begin{subequations}
\begin{align}
&[H, Q^{\pm}] = [H, N_{\text{PF}}] = 0, \label{eq:1.1a}\\
&[N_{\text{PF}}, Q^{\pm}] = \pm Q^{\pm}, \label{eq:1.1b}
\end{align}
\end{subequations}
and some multilinear relations among $H$, $Q^{+}$ and $Q^{-}$, which depend on the formulations but are reduced to the ordinary $\mathscr{N}=2$ supersymmetry algebra when $p=1$.
The most important feature of parasupersymmetric quantum mechanics of order $p$ is that the energy spectrum exhibits $(p+1)$-fold degeneracy.
This can be easily seen from the commutation relation \eqref{eq:1.1b} and the nilpotency $(Q^{\pm})^{p+1} = 0$: Since $Q^{+}$ and $Q^{-}$ raise and lower the parafermion number by one, by defining a boson state as an annihilated state by $Q^{-}$, one immediately sees that successive multiplications of $Q^{+}$ to the boson state yield $p$ distinct parafermion states with the same energy eigenvalue.

The purpose of this paper is to explore parasupersymmetry structures hidden behind degenerate spectra in single-particle quantum mechanics on graphs (quantum graphs).
In order to realize parasupersymmetric quantum mechanics of order $p$, we need to have some symmetry transformation which plays the role of parafermion number operator, or grading operator.
Such symmetry transformation must commute with the Hamiltonian and have $p+1$ distinct eigenvalues, because $N_{\text{PF}}$ splits the total Hilbert space into $p+1$ distinct subspaces.
A typical example of such symmetry transformations on graphs is $\mathbb{Z}_{p+1}$ cyclic rotation of $p+1$ edges, $\mathcal{Z}: \psi(x_{j}) \mapsto \psi(x_{j+1}) \pmod{p+1}$, whose eigenvalues are $\{1,q,\cdots,q^{p}\}$ with $q$ being the $(p+1)$th root of unity.
(Here $\psi(x_{j})$ is a wavefunction on the $j$th edge.)
If the Hamiltonian $H$ is independent of the edges, $H$ and $\mathcal{Z}$ obviously commute.
Thus the cyclic rotation $\mathcal{Z}$ is a good candidate for the grading operator $q^{N_{\text{PF}}}$, up to the question whether the boundary conditions are invariant under $\mathcal{Z}$.
Notice that the cyclic rotation $\mathcal{Z} = q^{N_{\text{PF}}}$ is a natural generalization of ordinary parity operator $\mathcal{P}: x \mapsto -x$, which plays the role of fermion parity $(-1)^{F}$ in one-dimensional quantum mechanics on $\mathbb{R}$ with hidden $\mathscr{N}=2$ supersymmetry \cite{Jakubsky:2010ki}.
Note also that the cyclic rotation of edges is well-defined only for equilateral graphs.

The rest of this paper is organized as follows.
In Section \ref{sec:2} we set up our model.
In this paper we consider compact equilateral graphs that consist of two vertices connected by $N$ edges of equal finite length.
For the sake of simplicity in this paper we focus on the free particle Hamiltonian (or minus the Laplace operator), which is obviously edge independent.
We then impose a generic $\mathbb{Z}_{N}$ cyclic symmetry in this system and obtain a $\mathbb{Z}_{N}$-graded Hilbert space.
In Section \ref{sec:3.1} we show that the Huang-Su parasupersymmetry algebra of order $p=N-1$ \cite{Huang:2010zzc} is hidden behind the spectral problem with $N$-fold degeneracy.
We classify boundary conditions invariant under parasupersymmetry transformations in Section \ref{sec:3.2} and then explicitly derive the parasuperspectrum in Section \ref{sec:4}.
It turns out that there are only four types of parasupersymmetry invariant boundary conditions, of which two leads to spontaneous breaking of parasupersymmetry.
Section \ref{sec:5} is devoted to conclusions and discussions.

\begin{figure}[t]
\centering
\subfigure[Unfolding picture.]{\input{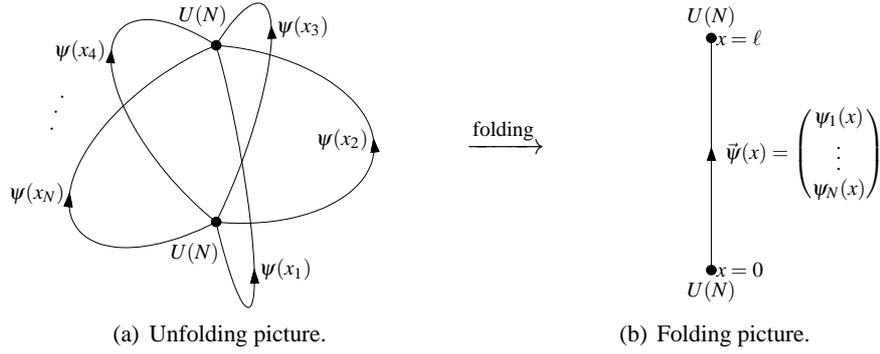} \label{fig:1a}}
\hspace{1cm}
\raisebox{2cm}{$\xrightarrow[]{\text{folding}}$}
\hspace{0cm}
\subfigure[Folding picture.]{
\xdefinecolor{rgb_000000}{rgb}{0,0,0}%
\setlength{\unitlength}{1cm}%
\begin{picture}(4,4)(0,0)%
\path(2,0.461538)(2,3.53846)
\path(2,1.92308)(2,2.07692)
\path(1.94728,1.91877)(2.05272,1.91877)
\path(1.95167,1.93195)(2.04833,1.93195)
\path(1.95607,1.94513)(2.04393,1.94513)
\path(1.96046,1.95831)(2.03954,1.95831)
\path(1.96485,1.97149)(2.03515,1.97149)
\path(1.96925,1.98466)(2.03075,1.98466)
\path(1.97364,1.99784)(2.02636,1.99784)
\path(1.97803,2.01102)(2.02197,2.01102)
\path(1.98243,2.0242)(2.01757,2.0242)
\path(1.98682,2.03738)(2.01318,2.03738)
\path(1.99121,2.05056)(2.00879,2.05056)
\path(1.99561,2.06374)(2.00439,2.06374)
\path(2.03954,1.91877)(2.03954,1.95831)
\path(2.02636,1.91877)(2.02636,1.99784)
\path(2.01318,1.91877)(2.01318,2.03738)
\path(2,1.91877)(2,2.07692)
\path(1.98682,1.91877)(1.98682,2.03738)
\path(1.97364,1.91877)(1.97364,1.99784)
\path(1.96046,1.91877)(1.96046,1.95831)
\path(2,1.91877)(2.05272,1.91877)(2,2.07692)(1.94728,1.91877)(2,1.91877)
\put(2,0.461538){\color{rgb_000000}$\allinethickness{0.070292cm}\circle{0.070292}$}%
\put(2,3.53846){\color{rgb_000000}$\allinethickness{0.070292cm}\circle{0.070292}$}%
\put(2.07029,0.461538){\makebox(0,0)[l]{\hbox{\color{rgb_000000}\scriptsize $x=0$}}}
\put(2.07029,3.53846){\makebox(0,0)[l]{\hbox{\color{rgb_000000}\scriptsize $x=\ell$}}}
\put(2,0.320955){\makebox(0,0)[t]{\hbox{\color{rgb_000000}\scriptsize $U(N)$}}}
\put(2,3.67905){\makebox(0,0)[b]{\hbox{\color{rgb_000000}\scriptsize $U(N)$}}}
\put(2.17573,2){\makebox(0,0)[l]{\hbox{\color{rgb_000000}\scriptsize $\Vec{\psi}(x) = \begin{pmatrix}\psi_{1}(x)\\ \vdots \\ \psi_{N}(x)\end{pmatrix}$}}}
\end{picture}%
 \label{fig:1b}}
\caption{(a) Carambola graph $\mathcal{C}_{N}$ and (b) its folding. Arrows indicate the direction of coordinates.}
\label{fig:1}
\end{figure}

\section{\texorpdfstring{$\mathbb{Z}_{N}$-graded Hilbert space}{ZN-graded Hilbert space}} \label{sec:2}
In this section we study non-relativistic quantum mechanics for a free particle on compact graphs with $\mathbb{Z}_{N}$ symmetry and its grading structure.
As mentioned in the previous section, in this paper we will focus on equilateral graphs that consist of two vertices connected by $N$ edges of the same length $\ell$.
Following Ref.\cite{Weigert:1998} we will call this graph a carambola graph and denote it by $\mathcal{C}_{N}$.\footnote{Carambola (also known as star fruit) is a golden-yellow tropical fruit with a star-shaped cross section.}
An important point to note is that, by folding the edges, quantum mechanical system on such graphs has an alternative equivalent description as a vector quantum mechanics on a single interval $(0, \ell)$; see Figure \ref{fig:1}.
This is referred to as the folding trick in the boundary conformal field theory literature \cite{Oshikawa:1996,Bachas:2001vj}.
As shown in Figure \ref{fig:1b}, the wavefunction of the system is given by an $N$-component complex column vector in the folding picture, $\Vec{\psi}(x) = (\psi_{1}(x), \cdots, \psi_{N}(x))^{T}$, where $T$ stands for transposition.
Namely, the Hilbert space is given by a tensor product
\begin{align}
\mathcal{H} = L^{2}(0, \ell) \otimes \mathbb{C}^{N}. \label{eq:2.1}
\end{align}
A free particle on the graph is described by the time-independent Schr\"odinger equation $H\Vec{\psi}(x) = E\Vec{\psi}(x)$, where $H$ is the Hamiltonian operator given by the $N \times N$ diagonal matrix
\begin{align}
H = -\frac{d^{2}}{dx^{2}} \otimes I, \label{eq:2.2}
\end{align}
with $I$ being the $N \times N$ identity matrix.
(Throughout this paper we will work in the units $\hbar = 2m = 1$.)
Given an observable, we have to specify the domain on which it becomes the self-adjoint operator.
For the case of the Hamiltonian, this can be done by specifying boundary conditions for $\Vec{\psi}$ consistent with the probability current conservations at each boundary (or Kirchhoff's law of probability current at the vertices).
It is well-known that probability current conservations lead to the following $U(N) \times U(N)$ family of boundary conditions (see e.g. Ref.\cite{Kuchment:2008}):
\begin{subequations}
\begin{align}
&(I + U)\Vec{\psi}^{\prime}(0) + iM_{0}(I - U)\Vec{\psi}(0) = \Vec{0}, \quad U \in U(N), \label{eq:2.3a}\\
&(I + \Tilde{U})\Vec{\psi}^{\prime}(\ell) - iM_{0}(I - \Tilde{U})\Vec{\psi}(\ell) = \Vec{0}, \quad \Tilde{U} \in U(N), \label{eq:2.3b}
\end{align}
\end{subequations}
where prime (${}^{\prime}$) indicates the derivative with respect to $x$.
$M_{0}$ is an arbitrary scale parameter with length dimension $-1$, which must be introduced to adjust the scaling dimensions of the first and second terms in the boundary conditions.
We note that without any loss of generality $M_{0}$ can be taken to be positive, $M_{0} > 0$.
The sign difference in the second terms of \eqref{eq:2.3a} and \eqref{eq:2.3b} is just for later convenience.

Let us next consider a $\mathbb{Z}_{N}$ symmetry transformation of edges that plays the role of the grading operator $(\mathrm{e}^{2\pi i/N})^{N_{\text{PF}}}$.
As mentioned in the previous section, a typical example of such symmetry transformations is a cyclic rotation of edges, $\psi_{j} \mapsto \psi_{j+1} \pmod{N}$.
However, it is not necessarily to be the cyclic rotation.
In this paper we consider a generic $\mathbb{Z}_{N}$ symmetry transformation given by the following unitary transformation:
\begin{align}
\mathcal{Z}: \Vec{\psi}(x) \mapsto (\mathcal{Z}\Vec{\psi})(x) = Z\Vec{\psi}(x), \label{eq:2.4}
\end{align}
where $Z \in U(N)$ is an $N \times N$ traceless unitary matrix satisfying $Z^{N} = I$ and $Z^{n} \neq I$ ($n = 1,\cdots,N-1$).
$Z$ has the eigenvalues $\{1,q,\cdots,q^{N-1}\}$ with $q = \exp(2\pi i/N)$ being the $N$th root of unity\footnote{In general, $q$ can be any primitive $N$th root of unity $q = \exp(2\pi ik/N)$, where integers $k$ and $N$ do not have common divisors. All the results of the paper are remained unchanged under this replacement.} such that it has the following spectral decomposition:
\begin{align}
Z = \sum_{n=0}^{N-1}q^{n}P_{n}, \label{eq:2.5}
\end{align}
where $P_{n}$ is a hermitian projection operator onto the $n$th eigenspace.
It should be noted that, in terms of the unitary matrix $Z$, such projection operator can be constructed as follows:
\begin{align}
P_{n} = \frac{I + q^{-n}Z + q^{-2n}Z^{2} + \cdots + q^{-(N-1)n}Z^{N-1}}{N}. \label{eq:2.6}
\end{align}
It is not difficult to show that thus constructed projection operator indeed satisfies the orthonormality relation $P_{n}P_{m} = \delta_{nm}P_{m}$, completeness relation $P_{0} + \cdots + P_{N-1} = I$, and hermiticity $P_{n}^{\dagger} = P_{n}$, where the first two relations follow from the identity $1 + q + \cdots + q^{N-1} = 0$.

Next we impose the system to be invariant (or symmetric) under the transformation $\mathcal{Z}$.
To this end, it is worthwhile to remember first the meaning of symmetry in quantum graphs, by following the discussion given by F\"ul\"op \textit{et al.} \cite{Fulop:2003} for the case of $U(2)$ family of point interactions on $S^{1}$.
The system is said to be symmetric under $\mathcal{Z}$ if the transformed state $\mathcal{Z}\Vec{\psi}$ satisfies i) the same Schr\"odinger equation and ii) the same boundary conditions as the original ones.
The first point is trivially satisfied: If $\Vec{\psi}$ satisfies the Schr\"odinger equation $H\Vec{\psi} = E\Vec{\psi}$, then the transformed state $\mathcal{Z}\Vec{\psi}$ also satisfies the same Schr\"odinger equation $H(\mathcal{Z}\Vec{\psi}) = E(\mathcal{Z}\Vec{\psi})$ because $H = -d^{2}/dx^{2} \otimes I$ and $\mathcal{Z} = 1 \otimes Z$ trivially commutes, $[H, \mathcal{Z}] = 0$.
The second point is, however, nontrivial: If $\Vec{\psi}$ satisfies \eqref{eq:2.3a} and \eqref{eq:2.3b}, then the boundary conditions for $\mathcal{Z}\Vec{\psi}$ become
\begin{subequations}
\begin{align}
&(I + ZUZ^{-1})(\mathcal{Z}\Vec{\psi})^{\prime}(0) + iM_{0}(I - ZUZ^{-1})(\mathcal{Z}\Vec{\psi})(0) = \Vec{0}, \label{eq:2.7a}\\
&(I + Z\Tilde{U}Z^{-1})(\mathcal{Z}\Vec{\psi})^{\prime}(\ell) - iM_{0}(I - Z\Tilde{U}Z^{-1})(\mathcal{Z}\Vec{\psi})(\ell) = \Vec{0}, \label{eq:2.7b}
\end{align}
\end{subequations}
which follow from multiplying $Z$ to Eqs. \eqref{eq:2.3a} and \eqref{eq:2.3b} from the left.
Comparing these to the original boundary conditions \eqref{eq:2.3a} and \eqref{eq:2.3b}, we see that $\mathcal{Z}$ induces the maps
\begin{align}
U \stackrel{\mathcal{Z}}{\mapsto} ZUZ^{-1}
\quad\text{and}\quad
\Tilde{U} \stackrel{\mathcal{Z}}{\mapsto} Z\Tilde{U}Z^{-1}. \label{eq:2.8}
\end{align}
If $U = ZUZ^{-1}$ and $\Tilde{U} = Z\Tilde{U}Z^{-1}$ (i.e., $[Z, U] = [Z, \Tilde{U}] = 0$), the boundary conditions are invariant under $\mathcal{Z}$.
To summarize, if $\mathcal{Z}$ is a symmetry of the system, the unitary matrices $Z$, $U$ and $\Tilde{U}$ must be all simultaneously diagonalizable; in other words, they have to share the same projection operators $P_{n}$.
Such unitary matrices enjoy the following parametric representations:
\begin{subequations}
\begin{align}
U
&= 	\sum_{n=0}^{N-1}\exp(i\theta_{n})P_{n}, \label{eq:2.9a}\\
\Tilde{U}
&= 	\sum_{n=0}^{N-1}\exp(i\Tilde{\theta}_{n})P_{n}, \label{eq:2.9b}
\end{align}
\end{subequations}
where $\theta_{n}$ and $\Tilde{\theta}_{n}$ ($n=0,1,\cdots,N-1$) are $2N$ independent parameters.
Eqs. \eqref{eq:2.9a} and \eqref{eq:2.9b} are the general solutions to the conditions $[Z, U] = [Z, \Tilde{U}] = 0$.

Now, both the Hamiltonian and boundary conditions commute with $\mathcal{Z}$, the Hilbert space completely splits into $N$ orthogonal subspaces; that is, $\mathcal{H}$ can be written as the following direct sum:
\begin{align}
\mathcal{H} = \mathcal{H}_{0} \oplus \mathcal{H}_{1} \oplus \cdots \oplus \mathcal{H}_{N-1}. \label{eq:2.10}
\end{align}
where $\mathcal{H}_{n} = \{\Vec{\psi}_{n} = P_{n}\Vec{\psi} \mid \Vec{\psi} \in \mathcal{H}\}$.
Notice that $\Vec{\psi}_{n} \in \mathcal{H}_{n}$ satisfies the eigenvalue equation $\mathcal{Z}\Vec{\psi}_{n} = q^{n}\Vec{\psi}_{n}$.
Let us next investigate the boundary conditions for $\Vec{\psi}_{n}$.
To this end, let $\Vec{e}_{n}$ be a simultaneous eigenvector of $Z$, $U$ and $\Tilde{U}$ that satisfy the orthonormality and completeness relations, $\Vec{e}_{n}^{\dagger}\Vec{e}_{m} = \delta_{nm}$ and $\Vec{e}_{0}\Vec{e}_{0}^{\dagger} + \cdots + \Vec{e}_{N-1}\Vec{e}_{N-1}^{\dagger} = I$.
The projection operator $P_{n}$ can then be written as $P_{n} = \Vec{e}_{n}\Vec{e}_{n}^{\dagger}$ such that $\Vec{\psi}_{n} = P_{n}\Vec{\psi}$ is cast into the following form:
\begin{align}
\Vec{\psi}_{n}(x) = \psi_{n}(x)\Vec{e}_{n}. \label{eq:2.11}
\end{align}
where $\psi_{n} = \Vec{e}_{n}^{\dagger}\Vec{\psi}$.
In what follows we will call the set of complete orthonormal eigenvectors $\{\Vec{e}_{0}, \cdots, \Vec{e}_{N-1}\}$ the (ordered) basis and $\psi_{n}(x)$ the component.
By multiplying $\Vec{e}_{n}^{\dagger}$ from the left, the boundary conditions \eqref{eq:2.3a} and \eqref{eq:2.3b} boil down to the following $2N$ independent Robin boundary conditions for the component $\psi_{n}$ ($n=0,1,\cdots,N-1$):
\begin{subequations}
\begin{align}
&\cos\left(\frac{\theta_{n}}{2}\right)\psi_{n}^{\prime}(0) + M_{0}\sin\left(\frac{\theta_{n}}{2}\right)\psi_{n}(0) = 0, \label{eq:2.12a}\\
&\cos\left(\frac{\Tilde{\theta}_{n}}{2}\right)\psi_{n}^{\prime}(\ell) - M_{0}\sin\left(\frac{\Tilde{\theta}_{n}}{2}\right)\psi_{n}(\ell) = 0, \label{eq:2.12b}
\end{align}
\end{subequations}
which follow from the eigenvalue equations for $\Vec{e}_{n}^{\dagger}$, $\Vec{e}_{n}^{\dagger}U = \mathrm{e}^{i\theta_{n}}\Vec{e}_{n}^{\dagger}$ and $\Vec{e}_{n}^{\dagger}\Tilde{U} = \mathrm{e}^{i\Tilde{\theta}_{n}}\Vec{e}_{n}^{\dagger}$.

To summarize, by imposing the $\mathbb{Z}_{N}$-symmetry \eqref{eq:2.4}, we have obtained the $\mathbb{Z}_{N}$-graded Hilbert space \eqref{eq:2.10} characterized by the eigenvalue $q^{n} = (\mathrm{e}^{2\pi i/N})^{n}$ of the unitary matrix $Z$.
In parasupersymmetric quantum mechanics language, this eigenvalue corresponds to the exponential of the parafermion number: $\mathcal{Z}$ counts the parafermion number modulo $N$.
Each subspace $\mathcal{H}_{n}$ carries the parafermion number $n$, and specified by the Robin boundary conditions \eqref{eq:2.12a} and \eqref{eq:2.12b}.

A few more comments are in order about two other symmetry transformations on the graph $\mathcal{C}_{N}$.
\begin{itemize}
\item The first symmetry transformation is a reflection $\mathcal{R}$ around the midpoint $x = \ell/2$, $\mathcal{R}: x \mapsto \ell - x$, which just flips the coordinate direction.
The action of $\mathcal{R}$ on $\Vec{\psi} \in \mathcal{H}$ is defined as
\begin{align}
\mathcal{R}: \Vec{\psi}(x) \mapsto (\mathcal{R}\Vec{\psi})(x) = \Vec{\psi}(\ell - x). \label{eq:2.13}
\end{align}
Obviously this $\mathbb{Z}_{2}$-transformation $\mathcal{R}$ commutes with the free Hamiltonian $H$ such that it preserves the energy spectrum.
However, $\mathcal{R}$ does not preserve the domain of $H$.
To see this, suppose that $\Vec{\psi}$ fulfills the boundary conditions.
Then, the transformed state $\mathcal{R}\Vec{\psi}$ satisfies
\begin{subequations}
\begin{align}
&-(I+U)(\mathcal{R}\Vec{\psi})^{\prime}(\ell) + iM_{0}(I-U)(\mathcal{R}\Vec{\psi})(\ell) = \Vec{0}, \label{eq:2.14a}\\
&-(I+\Tilde{U})(\mathcal{R}\Vec{\psi})^{\prime}(0) - iM_{0}(I - \Tilde{U})(\mathcal{R}\Vec{\psi})(0) = \Vec{0}, \label{eq:2.14b}
\end{align}
\end{subequations}
where we have used $(\mathcal{R}\Vec{\psi})^{\prime}(x) = -\Vec{\psi}^{\prime}(\ell - x)$.
Thus $\mathcal{R}$ induces the maps
\begin{align}
U \stackrel{\mathcal{R}}{\mapsto} \Tilde{U}
\quad\text{and}\quad
\Tilde{U} \stackrel{\mathcal{R}}{\mapsto} U. \label{eq:2.15}
\end{align}
If $U = \Tilde{U}$, the system is said to be $\mathcal{R}$-invariant.
If two systems are related by the maps, we say such systems are $\mathcal{R}$-dual.
As mentioned above, $\mathcal{R}$ is the spectrum-preserving transformation such that $\mathcal{R}$-dual systems are isospectral.
We will encounter this duality in Section \ref{sec:4.3}.

\item Given the $\mathbb{Z}_{N}$-transformation $\mathcal{Z}$, we can introduce another $\mathbb{Z}_{N}$-transformation $\mathcal{X}: \Vec{\psi}(x) \mapsto (\mathcal{X}\Vec{\psi})(x)$ that satisfies the $q$-commutation relation
\begin{align}
\mathcal{Z}\mathcal{X} = q\mathcal{X}\mathcal{Z}. \label{eq:2.16}
\end{align}
Indeed, $\mathcal{X}$ can be defined as $(\mathcal{X}\Vec{\psi})(x) = X\Vec{\psi}(x)$ with $X$ being an $N \times N$ traceless unitary matrix given by $X = \Vec{e}_{0}\Vec{e}_{N-1}^{\dagger} + \Vec{e}_{1}\Vec{e}_{0}^{\dagger} + \Vec{e}_{2}\Vec{e}_{1}^{\dagger} + \cdots + \Vec{e}_{N-1}\Vec{e}_{N-2}^{\dagger}$.
Thus constructed unitary matrix satisfies $X^{N} = I$ and $q$-commutation relation $ZX = qXZ$, so does $\mathcal{X}$.
A pair of unitary matrices $(X, Z)$ is called a Weyl pair and has been vastly studied both in mathematics and physics (see for review Ref.\cite{Kibler:2009}).\footnote{There are several conventions for the definition of a Weyl pair. In Ref.\cite{Kibler:2009} a pair $(X^{\dagger}, Z)$ is called  a Weyl pair.}
In the basis $\{\Vec{e}_{0}, \cdots, \Vec{e}_{N-1}\}$ the Weyl pair takes the following standard forms:
\begin{align}
X
= 	\begin{pmatrix}
	0 		& 0 		& \cdots 	& 0 		& 1\\
	1 		& 0 		& \cdots 	& 0 		& 0\\
	0 		& 1 		& \cdots 	& 0 		& 0\\
	\vdots 	& \vdots 	& \ddots 	& \vdots 	& \vdots\\
	0 		& 0 		& \cdots 	& 1 		& 0
	\end{pmatrix},
\quad
Z
= 	\begin{pmatrix}
	1 		& 0 		& 0 		& \cdots 	& 0\\
	0 		& q 		& 0 		& \cdots 	& 0\\
	0 		& 0 		& q^{2} 	& \cdots 	& 0\\
	\vdots 	& \vdots 	& \vdots 	& \ddots 	& \vdots\\
	0 		& 0 		& 0 		& \cdots 	& q^{N-1}
	\end{pmatrix}. \label{eq:2.17}
\end{align}
By construction it is obvious that $X$ and $X^{\dagger}$ shift the parafermion number $n$ by $\pm 1$
\begin{align}
X\Vec{e}_{n} = \Vec{e}_{n+1}
\quad\text{and}\quad
X^{\dagger}\Vec{e}_{n} = \Vec{e}_{n-1} \pmod{N}. \label{eq:2.18}
\end{align}
As we will see in Section \ref{sec:3.1}, the unitary matrix $X$ enables us to construct the parasupercharges $Q^{\pm}: \mathcal{H}_{n} \to \mathcal{H}_{n\pm1}$ in the basis independent way.
\end{itemize}

\section{\texorpdfstring{$\mathscr{N}=2$ parasupersymmetry of order $p=N-1$}{N=2 parasupersymmetry of order p=N-1}} \label{sec:3}
Roughly speaking, parasupersymmetric quantum mechanics of order $p$ is just a set of $p$ ordinary $\mathscr{N}=2$ supersymmetric quantum mechanical systems glued together in an intertwined way.
The point is a hierarchy of isospectral Hamiltonians based on the factorization method.
In this section we first try to construct the system to be $N$-fold degenerate by using the factorization method, and then reveal the underlying parasupersymmetry structure.

There are two important observations here.
The first is that, except for the case $\theta_{n}, \Tilde{\theta}_{n} \neq \pi \pmod{2\pi}$, the boundary conditions \eqref{eq:2.12a} and \eqref{eq:2.12b} can be compactly written as $(A_{\theta_{n}}^{+}\psi_{n})(0) = 0 = (A_{\Bar{\theta}_{n}}^{-}\psi_{n})(\ell)$, where $A_{\theta}^{\pm}$ are one-parameter family of first-order differential operators defined by
\begin{align}
A^{\pm}_{\theta} = \pm\frac{d}{dx} + M(\theta) \quad\text{with}\quad M(\theta) = M_{0}\tan\left(\frac{\theta}{2}\right). \label{eq:3.1}
\end{align}
Notice that $A^{-}_{\theta} = -A^{+}_{-\theta}$.
Note also that $A^{-}_{\theta}$ is a formal adjoint of $A^{+}_{\theta}$.

The second is that the free Hamiltonian $-d^{2}/dx^{2}$ can be factorized as $A^{\mp}_{\theta}A^{\pm}_{\theta} - M^{2}(\theta)$ for any $\theta$, which is the heart of hidden parasupersymmetry structures in our model.
To be more specific, let $H_{n} = -d^{2}/dx^{2}$ be the Hamiltonian operator for the component $\psi_{n}$ that satisfies $H_{n}\psi_{n} = E\psi_{n}$.
Then the above factorization enables us to construct the following hierarchy of Hamiltonians:
\begin{alignat}{7}
&H_{0}&\,\,=\,\,&A^{-}_{\alpha_{1}}A^{+}_{\alpha_{1}} - M^{2}(\alpha_{1})&\nonumber\\
&H_{1}&\,\,=\,\,&A^{+}_{\alpha_{1}}A^{-}_{\alpha_{1}} - M^{2}(\alpha_{1})&
\,\,=\,\,&A^{-}_{\alpha_{2}}A^{+}_{\alpha_{2}} - M^{2}(\alpha_{2})&\nonumber\\
&H_{2}&\,\,=\,\,& &
\,\,=\,\,&A^{+}_{\alpha_{2}}A^{-}_{\alpha_{2}} - M^{2}(\alpha_{2})&
\,\,=\,\,&A^{-}_{\alpha_{3}}A^{+}_{\alpha_{3}} - M^{2}(\alpha_{3})&\nonumber\\
&\,\,\vdots& & &
& &
\vdots\,\,\,\,& &\nonumber\\
&H_{N-1}&\,\,=\,\,& &
& &
\,\,=\,\,&A^{+}_{\alpha_{N-1}}A^{-}_{\alpha_{N-1}} - M^{2}(\alpha_{N-1}).&\nonumber
\end{alignat}
We emphasize that the parameters $\{\alpha_{n}\}_{n=1}^{N-1}$ are independent of the boundary condition parameters $\{\theta_{n}, \Tilde{\theta}_{n}\}_{n=0}^{N-1}$ at this stage.
Parasupersymmetry relations corresponding to this hierarchy of Hamiltonians are as follows:
\begin{subequations}
\begin{align}
&
\begin{cases}
\displaystyle
A^{+}_{\alpha_{1}}\psi_{0}(x) = \sqrt{E + M^{2}(\alpha_{1})}\psi_{1}(x), \\[1ex]
\displaystyle
A^{-}_{\alpha_{1}}\psi_{1}(x) = \sqrt{E + M^{2}(\alpha_{1})}\psi_{0}(x),
\end{cases} \label{eq:3.2a}\\
&
\begin{cases}
\displaystyle
A^{+}_{\alpha_{2}}\psi_{1}(x) = \sqrt{E + M^{2}(\alpha_{2})}\psi_{2}(x), \\[1ex]
\displaystyle
A^{-}_{\alpha_{2}}\psi_{2}(x) = \sqrt{E + M^{2}(\alpha_{2})}\psi_{1}(x),
\end{cases} \label{eq:3.2b}\\
&\hspace{3em}\vdots\nonumber\\
&
\begin{cases}
\displaystyle
A^{+}_{\alpha_{N-1}}\psi_{N-2}(x) = \sqrt{E + M^{2}(\alpha_{N-1})}\psi_{N-1}(x),\\[1ex]
\displaystyle
A^{-}_{\alpha_{N-1}}\psi_{N-1}(x) = \sqrt{E + M^{2}(\alpha_{N-1})}\psi_{N-2}(x).
\end{cases} \label{eq:3.2c}
\end{align}
\end{subequations}
Now our task is to tune the parameters $\{\theta_{n}, \Tilde{\theta}_{n}\}$ to be consistent with Eqs. \eqref{eq:3.2a}--\eqref{eq:3.2c}.
For example, if we impose the Robin boundary condition $(A^{+}_{\alpha_{1}}\psi_{0})(0) = 0$, which corresponds to the choice $\theta_{0} = \alpha_{1} \neq \pi \pmod{2\pi}$, the parasupersymmetry relation \eqref{eq:3.2a} implies $\psi_{1}$ must obey the Dirichlet boundary condition $\psi_{1}(0) = 0$, which corresponds to the choice $\theta_{1} = \pi \pmod{2\pi}$.
It follows from the relation \eqref{eq:3.2b} that the Dirichlet boundary condition $\psi_{1}(0) = 0$ leads to the Robin boundary condition for $\psi_{2}$, $(A^{-}_{\alpha_{2}}\psi_{2})(0) = 0$, which corresponds to the choice $\theta_{2} = -\alpha_{2} \neq \pi \pmod{2\pi}$.
This procedure can be easily extended for all $\psi_{n}$.
Before completing this program, however, let us check the hidden parasupersymmetry at the algebra level.

\subsection{Parasupersymmetry algebra} \label{sec:3.1}
The above constructed hierarchy of Hamiltonians based on the factorization $-d^{2}/dx^{2} = A^{\mp}_{\alpha}A^{\pm}_{\alpha} - M^{2}(\alpha)$ is well-described by the Huang-Su parasupersymmetry algebra of order $p = N-1$ \cite{Huang:2010zzc}.
To see this, let us first work in the basis $\{\Vec{e}_{0}, \Vec{e}_{1}, \cdots, \Vec{e}_{N-1}\}$ in which both $H$ and $Z$ become diagonal, $H = \mathrm{diag}(H_{0}, H_{1}, \cdots, H_{N-1})$ and $Z = \mathrm{diag}(1, q, \cdots, q^{N-1})$.
In this basis we define parasupercharges $Q^{\pm}$ as the following standard forms:
\begin{align}
Q^{+}
= 	\begin{pmatrix}
	0 				& 0 				& \cdots 	& 0 	& 0\\
	A^{+}_{\alpha_{1}} 	& 0 				& \cdots 	& 0 	& 0\\
	0 				& A^{+}_{\alpha_{2}} 	& \cdots 	& 0 	& 0\\
	\vdots 			& \vdots 			& \ddots 	& \vdots 	& \vdots\\
	0 				& 0 				& \cdots 	& A^{+}_{\alpha_{N-1}} 	& 0
	\end{pmatrix}, \quad
Q^{-}
= 	\begin{pmatrix}
	0 		& A^{-}_{\alpha_{1}} 	& 0 				& \cdots 	& 0\\
	0 		& 0 				& A^{-}_{\alpha_{2}} 	& \cdots 	& 0\\
	\vdots 	& \vdots 			& \vdots 			& \ddots 	& \vdots\\
	0 		& 0 				& 0 				& \cdots 	& A^{-}_{\alpha_{N-1}}\\
	0 		& 0				& 0				& \cdots 	& 0
	\end{pmatrix}. \label{eq:3.3}
\end{align}
It is straightforward to show that a set of operators $\{H, \mathcal{Z}, Q^{\pm}\}$ satisfies the following $\mathscr{N}=2$ parasupersymmetry algebra of order $p = N-1$:
\begin{subequations}
\begin{align}
&\mathcal{Z}^{N} = \mathrm{Id}, \quad \mathcal{Z}^{\dagger} = \mathcal{Z}^{-1}, \label{eq:3.4a}\\
&(Q^{\pm})^{N} = 0, \label{eq:3.4b}\\
&\mathcal{Z}Q^{\pm} = q^{\pm1}Q^{\pm}\mathcal{Z}, \label{eq:3.4c}\\
&[H, Q^{\pm}] = [H, \mathcal{Z}] = 0, \label{eq:3.4d}\\
&\sum_{m=0}^{N-1}(Q^{+})^{N-1-m}(Q^{-})^{N-1}(Q^{+})^{m} = \prod_{n=1}^{N-1}\bigl[H + M^{2}(\alpha_{n})\bigr], \label{eq:3.4e}
\end{align}
\end{subequations}
where $\mathrm{Id}$ is the identity operator on $\mathcal{H}$.
We should emphasize that, as noted in Ref. \cite{Huang:2010zzc}, the parasupercharges \eqref{eq:3.3} satisfy the Rubakov-Spiridonov-Khare-Tomiya parasupersymmetry algebra \cite{Rubakov:1988mv,Khare:1992gy,Tomiya:1992} $\sum_{m=0}^{N-1}(Q^{+})^{N-1-m}Q^{-}(Q^{+})^{m} = (N-1)(Q^{+})^{N-2}H$ (and its hermitian conjugate) if the parameters $\{M(\alpha_{n})\}$ satisfy the condition $\sum_{n=1}^{N-1}M^{2}(\alpha_{n}) = 0$.
The Huang-Su multilinear relation \eqref{eq:3.4e}, one the other hand, remains valid for any $M(\alpha_{n})$.
We also emphasize that the case $N=2$ reduces to the ordinary $\mathscr{N} = 2$ supersymmetry algebra, $Q^{+}Q^{-} + Q^{-}Q^{+} = H + M^{2}(\alpha_{1})$, where the origin of energy is shifted by $M^{2}(\alpha_{1})$.

Let us next move on to an arbitrary basis.
Although the algebra \eqref{eq:3.4a}--\eqref{eq:3.4e} is independent of the choice of the basis, it is more desirable to construct the parasupercharges $Q^{\pm}$ in the basis independent way.
This can done by using the projection operator $P_{n}$ and the unitary matrix $X$ as follows:
\begin{subequations}
\begin{align}
Q^{+}
&= 	\sum_{n=0}^{N-2}A^{+}_{\alpha_{n+1}} \otimes XP_{n}, \label{eq:3.5a}\\
Q^{-}
&= 	\sum_{n=1}^{N-1}A^{-}_{\alpha_{n}} \otimes X^{\dagger}P_{n}. \label{eq:3.5b}
\end{align}
\end{subequations}
Interpretations of these expressions are obvious: When $Q^{\pm}$ act on $\Vec{\psi} = \Vec{\psi}_{0} + \Vec{\psi}_{1} + \cdots + \Vec{\psi}_{N-1} \in \mathcal{H}$, the projection operator $P_{n}$ picks up the mode $\Vec{\psi}_{n} = \psi_{n}\Vec{e}_{n}$.
Then $X$ shifts the basis vector $\Vec{e}_{n}$ to $\Vec{e}_{n+1}$ (see Eq. \eqref{eq:2.18}) and $A^{+}_{\alpha_{n+1}}$ shifts the component $\psi_{n}$ to $\psi_{n+1}$.
Similarly, $X^{\dagger}$ shifts the basis vector $\Vec{e}_{n}$ to $\Vec{e}_{n-1}$ and $A^{+}_{\alpha_{n}}$ shifts the component $\psi_{n}$ to $\psi_{n-1}$.
Thus we obtain the following complex (known as the ``paracomplex'' \cite{Picken:1990da,Stosic:2004}):
\begin{align}
0~\substack{\phantom{Q^{+}_{0}}\\ \phantom{\longrightarrow}\\ \longleftarrow\\ Q^{-}_{0}}~
\mathcal{H}_{0}~\substack{Q^{+}_{0}\\ \longrightarrow\\ \longleftarrow\\ Q^{-}_{1}}~
\mathcal{H}_{1}~\substack{Q^{+}_{1}\\ \longrightarrow\\ \longleftarrow\\ Q^{-}_{2}}~
\cdots
~\substack{Q^{+}_{N-2}\\ \longrightarrow\\ \longleftarrow\\ Q^{-}_{N-1}}~
\mathcal{H}_{N-1}~\substack{Q^{+}_{N-1}\\ \longrightarrow\\ \phantom{\longleftarrow}\\ \phantom{Q^{-}_{N-1}}}~0, \label{eq:3.6}
\end{align}
where $Q^{\pm}_{n} = Q^{\pm}P_{n}$.
Recall that the projection operators are written in terms of $Z$; see Eq. \eqref{eq:2.6}.
Hence, in order to construct the parasupercharges, we need only the Weyl pair $(X, Z)$ and the first-order differential operators $A^{\pm}_{\alpha_{n}}$.

\subsection{Parasupersymmetry invariant boundary conditions} \label{sec:3.2}
The boundary conditions are said to be parasupersymmetry invariant if the transformed states $Q^{\pm}\Vec{\psi}_{n}$ satisfy the same boundary conditions as those for $\Vec{\psi}_{n\pm1}$.
In this section we classify such parasupersymmetry invariant boundary conditions in our model.

To this end, we first note that, once we fix the boundary conditions for the bosonic sector $\mathcal{H}_{0} \ni \Vec{\psi}_{0}$, the boundary conditions for the parafermionic sectors $\mathcal{H}_{n}$ ($n=1,\cdots,N-1$) are automatically determined via parasupersymmetry relations \eqref{eq:3.2a}--\eqref{eq:3.2c}.
Note also that there are two possibilities for the bosonic sector: the Robin boundary condition $(A^{+}_{\alpha_{1}}\psi_{0})(0) = 0$ or Dirichlet boundary condition $\psi_{0}(0) = 0$, the former corresponds to the choice $\theta_{0} = \alpha_{1} \neq \pi \pmod{2\pi}$ and the latter $\theta_{0} = \pi \pmod{2\pi}$.
Thus there are two distinct types of boundary conditions at each boundary, one is the sequence of boundary conditions that starts from the Robin boundary condition for $\Vec{\psi}_{0}$ and the other the Dirichlet boundary condition.
Let us first study the former type.

\paragraph{Type R.}
If we impose the Robin boundary condition $(A^{+}_{\alpha_{1}}\psi_{0})(0) = 0$, then the parasupersymmetry relation \eqref{eq:3.2a} implies that $\psi_{1} \in \mathcal{H}_{1}$ must satisfies the Dirichlet boundary condition $\psi_{1}(0) = 0$.
From the parasupersymmetry relation \eqref{eq:3.2b}, the Dirichlet boundary condition for $\psi_{1}$ implies the Robin boundary conditions for $\psi_{2}$, $(A^{-}_{\alpha_{2}}\psi_{2})(0) = 0$.
An important point to note is that, if $\alpha_{2} = -\alpha_{3}$, the boundary conditions $(A^{-}_{\alpha_{2}}\psi_{2})(0) = 0$ and $(A^{+}_{\alpha_{3}}\psi_{2})(0) = 0$ are compatible with each other, because $A^{-}_{\alpha} = -A^{+}_{-\alpha}$ for any $\alpha$.
With this choice of the parameter, it follows from the parasupersymmetry relation between $\psi_{2}$ and $\psi_{3}$ that the Robin boundary condition $(A^{-}_{\alpha_{2}}\psi_{2})(0) = 0$ leads to the Dirichlet boundary condition for $\psi_{3}$, $\psi_{3}(0) = 0$.
This procedure is easily extended to all $\psi_{n} \in \mathcal{H}_{n}$.
Basically, the Robin and Dirichlet boundary conditions appear in alternating order.
The resultant parasupersymmetry invariant boundary conditions are summarized as follows:
\begin{subequations}
\begin{alignat}{2}
&\text{Case $N$ even:}~&
\begin{cases}
\displaystyle
(A^{+}_{\alpha_{n+1}}\psi_{n})(0) = 0 	& (n = 0,2,4,\cdots,N-2), \\[1ex]
\displaystyle
\psi_{n}(0) = 0 						& (n = 1,3,5,\cdots,N-1),
\end{cases}& \label{eq:3.7a}\\
&\text{Case $N$ odd:}&
\begin{cases}
\displaystyle
(A^{+}_{\alpha_{n+1}}\psi_{n})(0) = 0 	& (n = 0,2,4,\cdots,N-3), \\[1ex]
\displaystyle
\psi_{n}(0) = 0 						& (n = 1,3,5,\cdots,N-2), \\[1ex]
\displaystyle
(A^{-}_{\alpha_{n}}\psi_{n})(0) = 0 		& (n = N-1),
\end{cases}& \label{eq:3.7b}
\end{alignat}
\end{subequations}
where $\{\alpha_{n}\}_{n=1}^{N-1}$ are tuned to fulfill the constraints $\alpha_{n} = -\alpha_{n+1}$ with $n=2,4,6,\cdots,N-2$ for $N$ even and $n=2,4,6,\cdots,N-3$ for $N$ odd.
Hence, for $N$ even, we have $N/2$ independent parameters $\{\alpha_{1}, \alpha_{3}, \alpha_{5}, \cdots, \alpha_{N-1}\}$ in the boundary conditions and parasupercharges, while for $N$ odd we have $(N+1)/2$ independent parameters $\{\alpha_{1}, \alpha_{3}, \alpha_{5}, \cdots, \alpha_{N-2}, \alpha_{N-1}\}$.

\paragraph{Type D.}
Let us move on to the sequence that starts from the Dirichlet boundary condition for the bosonic sector, $\psi_{0}(0) = 0$.
The discussion is almost the same as the previous one.
The result is as follows:
\begin{subequations}
\begin{alignat}{2}
&\text{Case $N$ even:}~&
\begin{cases}
\displaystyle
\psi_{n}(0) = 0 					& (n=0,2,4,\cdots,N-2), \\[1ex]
\displaystyle
(A^{-}_{\alpha_{n}}\psi_{n})(0) = 0 	& (n=1,3,5,\cdots,N-1),
\end{cases}& \label{eq:3.8a}\\
&\text{Case $N$ odd:}&
\begin{cases}
\displaystyle
\psi_{n}(0) = 0 					& (n=0,2,4,\cdots,N-1), \\[1ex]
\displaystyle
(A^{-}_{\alpha_{n}}\psi_{n})(0) = 0 	& (n=1,3,5,\cdots,N-2).
\end{cases}& \label{eq:3.8b}
\end{alignat}
\end{subequations}
Here the parameters $\{\alpha_{n}\}_{n=1}^{N-1}$ must satisfy the constraints $\alpha_{n} = -\alpha_{n-1}$, where $n=2,4,6,\cdots,N-2$ for $N$ even and $n=2,4,6,\cdots,N-1$ for $N$ odd.
We thus have $N/2$ independent parameters $\{\alpha_{1}, \alpha_{3}, \alpha_{5}, \cdots, \alpha_{N-1}\}$ for $N$ even, while for $N$ odd we have $(N-1)/2$ independent parameters $\{\alpha_{1}, \alpha_{3}, \alpha_{5}, \cdots, \alpha_{N-2}\}$.

\vskip 1.6em

Parasupersymmetry invariant boundary conditions at $x = \ell$ are given by just replacing the argument $x = 0$ by $x = \ell$ in Eqs. \eqref{eq:3.7a}--\eqref{eq:3.8b}.
Having two types of boundary conditions at each boundary, we see that there are $2 \times 2 = 4$ distinct sets of boundary conditions consistent with the parasupersymmetry; namely, type (R, R), type (D, D), type (R, D) and type (D, R), where type (R, D) for example refers to a combination of type R boundary conditions at $x=0$ and type D boundary conditions at $x=\ell$.
It should be noted that, due to the parameter constraints $\alpha_{n} = -\alpha_{n+1}$ and $\alpha_{n} = -\alpha_{n-1}$ ($n=2,4,6,\cdots$), there appears only one independent parameter $\alpha := \alpha_{1}$ in the type (R, D) and (D, R) boundary conditions.

\section{Parasuperspectrum} \label{sec:4}
In this section we solve the Schr\"odinger equation $H\Vec{\psi}(x) = E\Vec{\psi}(x)$ with the parasupersymmetry invariant boundary conditions and then derive the spectrum.
We also discuss the spontaneous breaking of parasupersymmetry.

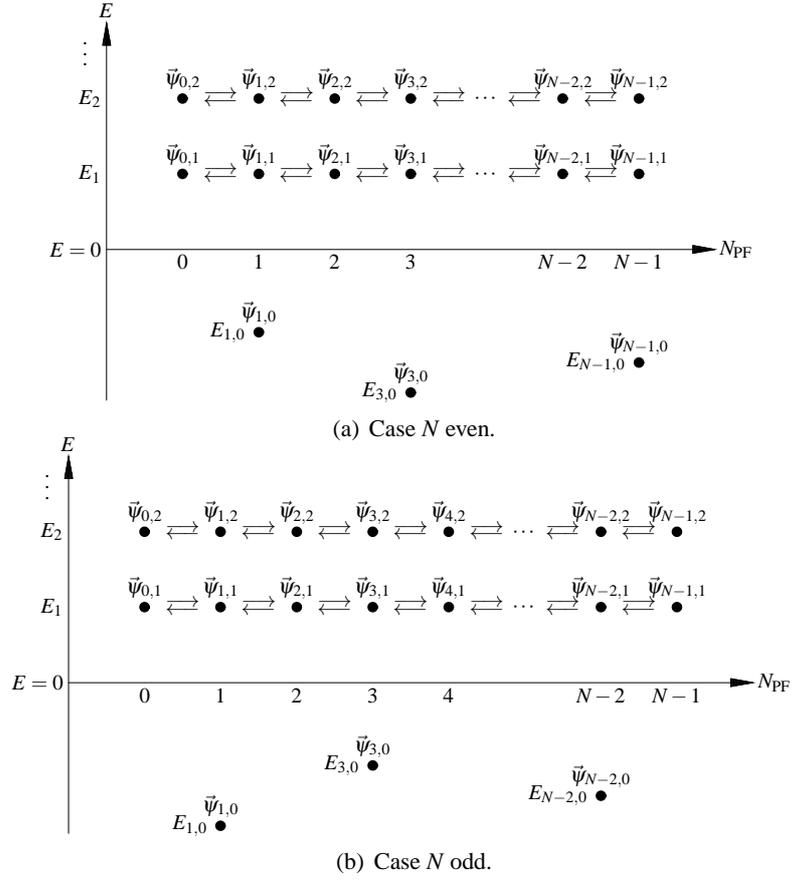
\begin{figure}[t]
\centering
\subfigure[Case $N$ even.]{
\xdefinecolor{rgb_000000}{rgb}{0,0,0}%
\setlength{\unitlength}{1cm}%
\begin{picture}(8,5)(0,0)%
\path(0,0)(0,5)
\path(-0.052719,4.71005)(0.052719,4.71005)
\path(-0.0502085,4.72385)(0.0502085,4.72385)
\path(-0.0476981,4.73766)(0.0476981,4.73766)
\path(-0.0451877,4.75147)(0.0451877,4.75147)
\path(-0.0426773,4.76528)(0.0426773,4.76528)
\path(-0.0401668,4.77908)(0.0401668,4.77908)
\path(-0.0376564,4.79289)(0.0376564,4.79289)
\path(-0.035146,4.8067)(0.035146,4.8067)
\path(-0.0326356,4.8205)(0.0326356,4.8205)
\path(-0.0301251,4.83431)(0.0301251,4.83431)
\path(-0.0276147,4.84812)(0.0276147,4.84812)
\path(-0.0251043,4.86193)(0.0251043,4.86193)
\path(-0.0225938,4.87573)(0.0225938,4.87573)
\path(-0.0200834,4.88954)(0.0200834,4.88954)
\path(-0.017573,4.90335)(0.017573,4.90335)
\path(-0.0150626,4.91716)(0.0150626,4.91716)
\path(-0.0125521,4.93096)(0.0125521,4.93096)
\path(-0.0100417,4.94477)(0.0100417,4.94477)
\path(-0.00753128,4.95858)(0.00753128,4.95858)
\path(-0.00502085,4.97239)(0.00502085,4.97239)
\path(-0.00251043,4.98619)(0.00251043,4.98619)
\path(0.0395392,4.71005)(0.0395392,4.78253)
\path(0.0263595,4.71005)(0.0263595,4.85502)
\path(0.0131797,4.71005)(0.0131797,4.92751)
\path(0,4.71005)(0,5)
\path(-0.0131797,4.71005)(-0.0131797,4.92751)
\path(-0.0263595,4.71005)(-0.0263595,4.85502)
\path(-0.0395392,4.71005)(-0.0395392,4.78253)
\path(0,4.71005)(0.052719,4.71005)(0,5)(-0.052719,4.71005)(0,4.71005)
\path(0,2)(8,2)
\path(7.98619,1.99749)(7.98409,2.00289)
\path(7.97239,1.99498)(7.96818,2.00578)
\path(7.95858,1.99247)(7.95228,2.00868)
\path(7.94477,1.98996)(7.93637,2.01157)
\path(7.93096,1.98745)(7.92046,2.01446)
\path(7.91716,1.98494)(7.90455,2.01735)
\path(7.90335,1.98243)(7.88864,2.02025)
\path(7.88954,1.97992)(7.87274,2.02314)
\path(7.87573,1.97741)(7.85683,2.02603)
\path(7.86193,1.9749)(7.84092,2.02892)
\path(7.84812,1.97239)(7.82501,2.03182)
\path(7.83431,1.96987)(7.8091,2.03471)
\path(7.8205,1.96736)(7.7932,2.0376)
\path(7.8067,1.96485)(7.77729,2.04049)
\path(7.79289,1.96234)(7.76138,2.04339)
\path(7.77908,1.95983)(7.74547,2.04628)
\path(7.76528,1.95732)(7.72956,2.04917)
\path(7.75147,1.95481)(7.71366,2.05206)
\path(7.73766,1.9523)(7.71005,2.02333)
\path(7.72385,1.94979)(7.71005,1.9853)
\path(7.73641,2.04793)(7.71005,2.03768)
\path(7.76276,2.04313)(7.71005,2.02264)
\path(7.78912,2.03834)(7.71005,2.0076)
\path(7.81548,2.03355)(7.71005,1.99255)
\path(7.84184,2.02876)(7.71005,1.97751)
\path(7.8682,2.02396)(7.71005,1.96247)
\path(7.89456,2.01917)(7.71005,1.94743)
\path(7.92092,2.01438)(7.78199,1.96036)
\path(7.94728,2.00959)(7.85466,1.97357)
\path(7.97364,2.00479)(7.92733,1.98679)
\path(7.71005,2)(7.71005,1.94728)(8,2)(7.71005,2.05272)(7.71005,2)
\put(1,3){\color{rgb_000000}$\allinethickness{0.070292cm}\circle{0.070292}$}%
\put(1,4){\color{rgb_000000}$\allinethickness{0.070292cm}\circle{0.070292}$}%
\put(2,3){\color{rgb_000000}$\allinethickness{0.070292cm}\circle{0.070292}$}%
\put(2,4){\color{rgb_000000}$\allinethickness{0.070292cm}\circle{0.070292}$}%
\put(3,3){\color{rgb_000000}$\allinethickness{0.070292cm}\circle{0.070292}$}%
\put(3,4){\color{rgb_000000}$\allinethickness{0.070292cm}\circle{0.070292}$}%
\put(4,3){\color{rgb_000000}$\allinethickness{0.070292cm}\circle{0.070292}$}%
\put(4,4){\color{rgb_000000}$\allinethickness{0.070292cm}\circle{0.070292}$}%
\put(5,3){\makebox(0,0)[c]{\hbox{\color{rgb_000000}\scriptsize $\cdots$}}}
\put(5,4){\makebox(0,0)[c]{\hbox{\color{rgb_000000}\scriptsize $\cdots$}}}
\put(6,3){\color{rgb_000000}$\allinethickness{0.070292cm}\circle{0.070292}$}%
\put(6,4){\color{rgb_000000}$\allinethickness{0.070292cm}\circle{0.070292}$}%
\put(7,3){\color{rgb_000000}$\allinethickness{0.070292cm}\circle{0.070292}$}%
\put(7,4){\color{rgb_000000}$\allinethickness{0.070292cm}\circle{0.070292}$}%
\put(2,0.9){\color{rgb_000000}$\allinethickness{0.070292cm}\circle{0.070292}$}%
\put(4,0.1){\color{rgb_000000}$\allinethickness{0.070292cm}\circle{0.070292}$}%
\put(7,0.5){\color{rgb_000000}$\allinethickness{0.070292cm}\circle{0.070292}$}%
\put(1.5,3){\makebox(0,0)[b]{\hbox{\color{rgb_000000}\scriptsize $\longrightarrow$}}}
\put(1.5,3){\makebox(0,0)[t]{\hbox{\color{rgb_000000}\scriptsize $\longleftarrow$}}}
\put(1.5,4){\makebox(0,0)[b]{\hbox{\color{rgb_000000}\scriptsize $\longrightarrow$}}}
\put(1.5,4){\makebox(0,0)[t]{\hbox{\color{rgb_000000}\scriptsize $\longleftarrow$}}}
\put(2.5,3){\makebox(0,0)[b]{\hbox{\color{rgb_000000}\scriptsize $\longrightarrow$}}}
\put(2.5,3){\makebox(0,0)[t]{\hbox{\color{rgb_000000}\scriptsize $\longleftarrow$}}}
\put(2.5,4){\makebox(0,0)[b]{\hbox{\color{rgb_000000}\scriptsize $\longrightarrow$}}}
\put(2.5,4){\makebox(0,0)[t]{\hbox{\color{rgb_000000}\scriptsize $\longleftarrow$}}}
\put(3.5,3){\makebox(0,0)[b]{\hbox{\color{rgb_000000}\scriptsize $\longrightarrow$}}}
\put(3.5,3){\makebox(0,0)[t]{\hbox{\color{rgb_000000}\scriptsize $\longleftarrow$}}}
\put(3.5,4){\makebox(0,0)[b]{\hbox{\color{rgb_000000}\scriptsize $\longrightarrow$}}}
\put(3.5,4){\makebox(0,0)[t]{\hbox{\color{rgb_000000}\scriptsize $\longleftarrow$}}}
\put(4.5,3){\makebox(0,0)[b]{\hbox{\color{rgb_000000}\scriptsize $\longrightarrow$}}}
\put(4.5,3){\makebox(0,0)[t]{\hbox{\color{rgb_000000}\scriptsize $\longleftarrow$}}}
\put(4.5,4){\makebox(0,0)[b]{\hbox{\color{rgb_000000}\scriptsize $\longrightarrow$}}}
\put(4.5,4){\makebox(0,0)[t]{\hbox{\color{rgb_000000}\scriptsize $\longleftarrow$}}}
\put(5.5,3){\makebox(0,0)[b]{\hbox{\color{rgb_000000}\scriptsize $\longrightarrow$}}}
\put(5.5,3){\makebox(0,0)[t]{\hbox{\color{rgb_000000}\scriptsize $\longleftarrow$}}}
\put(5.5,4){\makebox(0,0)[b]{\hbox{\color{rgb_000000}\scriptsize $\longrightarrow$}}}
\put(5.5,4){\makebox(0,0)[t]{\hbox{\color{rgb_000000}\scriptsize $\longleftarrow$}}}
\put(6.5,3){\makebox(0,0)[b]{\hbox{\color{rgb_000000}\scriptsize $\longrightarrow$}}}
\put(6.5,3){\makebox(0,0)[t]{\hbox{\color{rgb_000000}\scriptsize $\longleftarrow$}}}
\put(6.5,4){\makebox(0,0)[b]{\hbox{\color{rgb_000000}\scriptsize $\longrightarrow$}}}
\put(6.5,4){\makebox(0,0)[t]{\hbox{\color{rgb_000000}\scriptsize $\longleftarrow$}}}
\put(0,5.07029){\makebox(0,0)[b]{\hbox{\color{rgb_000000}\scriptsize $E$}}}
\put(8.07029,2){\makebox(0,0)[l]{\hbox{\color{rgb_000000}\scriptsize $N_{\text{PF}}$}}}
\put(-0.070292,2){\makebox(0,0)[r]{\hbox{\color{rgb_000000}\scriptsize $E=0$}}}
\put(-0.070292,3){\makebox(0,0)[r]{\hbox{\color{rgb_000000}\scriptsize $E_{1}$}}}
\put(-0.070292,4){\makebox(0,0)[r]{\hbox{\color{rgb_000000}\scriptsize $E_{2}$}}}
\put(-0.246022,4.7){\makebox(0,0)[r]{\hbox{\color{rgb_000000}\scriptsize $\vdots$}}}
\put(1,1.92971){\makebox(0,0)[t]{\hbox{\color{rgb_000000}\scriptsize $0$}}}
\put(2,1.92971){\makebox(0,0)[t]{\hbox{\color{rgb_000000}\scriptsize $1$}}}
\put(3,1.92971){\makebox(0,0)[t]{\hbox{\color{rgb_000000}\scriptsize $2$}}}
\put(4,1.92971){\makebox(0,0)[t]{\hbox{\color{rgb_000000}\scriptsize $3$}}}
\put(6,1.92971){\makebox(0,0)[t]{\hbox{\color{rgb_000000}\scriptsize $N-2$}}}
\put(7,1.92971){\makebox(0,0)[t]{\hbox{\color{rgb_000000}\scriptsize $N-1$}}}
\put(1,3.10544){\makebox(0,0)[b]{\hbox{\color{rgb_000000}\scriptsize $\Vec{\psi}_{0,1}$}}}
\put(2,3.10544){\makebox(0,0)[b]{\hbox{\color{rgb_000000}\scriptsize $\Vec{\psi}_{1,1}$}}}
\put(3,3.10544){\makebox(0,0)[b]{\hbox{\color{rgb_000000}\scriptsize $\Vec{\psi}_{2,1}$}}}
\put(4,3.10544){\makebox(0,0)[b]{\hbox{\color{rgb_000000}\scriptsize $\Vec{\psi}_{3,1}$}}}
\put(6,3.10544){\makebox(0,0)[b]{\hbox{\color{rgb_000000}\scriptsize $\Vec{\psi}_{N-2,1}$}}}
\put(7,3.10544){\makebox(0,0)[b]{\hbox{\color{rgb_000000}\scriptsize $\Vec{\psi}_{N-1,1}$}}}
\put(1,4.10544){\makebox(0,0)[b]{\hbox{\color{rgb_000000}\scriptsize $\Vec{\psi}_{0,2}$}}}
\put(2,4.10544){\makebox(0,0)[b]{\hbox{\color{rgb_000000}\scriptsize $\Vec{\psi}_{1,2}$}}}
\put(3,4.10544){\makebox(0,0)[b]{\hbox{\color{rgb_000000}\scriptsize $\Vec{\psi}_{2,2}$}}}
\put(4,4.10544){\makebox(0,0)[b]{\hbox{\color{rgb_000000}\scriptsize $\Vec{\psi}_{3,2}$}}}
\put(6,4.10544){\makebox(0,0)[b]{\hbox{\color{rgb_000000}\scriptsize $\Vec{\psi}_{N-2,2}$}}}
\put(7,4.10544){\makebox(0,0)[b]{\hbox{\color{rgb_000000}\scriptsize $\Vec{\psi}_{N-1,2}$}}}
\put(1.82427,0.9){\makebox(0,0)[r]{\hbox{\color{rgb_000000}\scriptsize $E_{1,0}$}}}
\put(3.82427,0.1){\makebox(0,0)[r]{\hbox{\color{rgb_000000}\scriptsize $E_{3,0}$}}}
\put(6.82427,0.5){\makebox(0,0)[r]{\hbox{\color{rgb_000000}\scriptsize $E_{N-1,0}$}}}
\put(2,1.00544){\makebox(0,0)[b]{\hbox{\color{rgb_000000}\scriptsize $\Vec{\psi}_{1,0}$}}}
\put(4,0.205438){\makebox(0,0)[b]{\hbox{\color{rgb_000000}\scriptsize $\Vec{\psi}_{3,0}$}}}
\put(7,0.605438){\makebox(0,0)[b]{\hbox{\color{rgb_000000}\scriptsize $\Vec{\psi}_{N-1,0}$}}}
\end{picture}%
 \label{fig:2a}} \hspace{1cm}
\subfigure[Case $N$ odd.]{
\xdefinecolor{rgb_000000}{rgb}{0,0,0}%
\setlength{\unitlength}{1cm}%
\begin{picture}(9,5)(0,0)%
\path(0,0)(0,5)
\path(-0.052719,4.71005)(0.052719,4.71005)
\path(-0.0502085,4.72385)(0.0502085,4.72385)
\path(-0.0476981,4.73766)(0.0476981,4.73766)
\path(-0.0451877,4.75147)(0.0451877,4.75147)
\path(-0.0426773,4.76528)(0.0426773,4.76528)
\path(-0.0401668,4.77908)(0.0401668,4.77908)
\path(-0.0376564,4.79289)(0.0376564,4.79289)
\path(-0.035146,4.8067)(0.035146,4.8067)
\path(-0.0326356,4.8205)(0.0326356,4.8205)
\path(-0.0301251,4.83431)(0.0301251,4.83431)
\path(-0.0276147,4.84812)(0.0276147,4.84812)
\path(-0.0251043,4.86193)(0.0251043,4.86193)
\path(-0.0225938,4.87573)(0.0225938,4.87573)
\path(-0.0200834,4.88954)(0.0200834,4.88954)
\path(-0.017573,4.90335)(0.017573,4.90335)
\path(-0.0150626,4.91716)(0.0150626,4.91716)
\path(-0.0125521,4.93096)(0.0125521,4.93096)
\path(-0.0100417,4.94477)(0.0100417,4.94477)
\path(-0.00753128,4.95858)(0.00753128,4.95858)
\path(-0.00502085,4.97239)(0.00502085,4.97239)
\path(-0.00251043,4.98619)(0.00251043,4.98619)
\path(0.0395392,4.71005)(0.0395392,4.78253)
\path(0.0263595,4.71005)(0.0263595,4.85502)
\path(0.0131797,4.71005)(0.0131797,4.92751)
\path(0,4.71005)(0,5)
\path(-0.0131797,4.71005)(-0.0131797,4.92751)
\path(-0.0263595,4.71005)(-0.0263595,4.85502)
\path(-0.0395392,4.71005)(-0.0395392,4.78253)
\path(0,4.71005)(0.052719,4.71005)(0,5)(-0.052719,4.71005)(0,4.71005)
\path(0,2)(9,2)
\path(8.98619,1.99749)(8.98409,2.00289)
\path(8.97239,1.99498)(8.96818,2.00578)
\path(8.95858,1.99247)(8.95228,2.00868)
\path(8.94477,1.98996)(8.93637,2.01157)
\path(8.93096,1.98745)(8.92046,2.01446)
\path(8.91716,1.98494)(8.90455,2.01735)
\path(8.90335,1.98243)(8.88864,2.02025)
\path(8.88954,1.97992)(8.87274,2.02314)
\path(8.87573,1.97741)(8.85683,2.02603)
\path(8.86193,1.9749)(8.84092,2.02892)
\path(8.84812,1.97239)(8.82501,2.03182)
\path(8.83431,1.96987)(8.8091,2.03471)
\path(8.8205,1.96736)(8.7932,2.0376)
\path(8.8067,1.96485)(8.77729,2.04049)
\path(8.79289,1.96234)(8.76138,2.04339)
\path(8.77908,1.95983)(8.74547,2.04628)
\path(8.76528,1.95732)(8.72956,2.04917)
\path(8.75147,1.95481)(8.71366,2.05206)
\path(8.73766,1.9523)(8.71005,2.02333)
\path(8.72385,1.94979)(8.71005,1.9853)
\path(8.73641,2.04793)(8.71005,2.03768)
\path(8.76276,2.04313)(8.71005,2.02264)
\path(8.78912,2.03834)(8.71005,2.0076)
\path(8.81548,2.03355)(8.71005,1.99255)
\path(8.84184,2.02876)(8.71005,1.97751)
\path(8.8682,2.02396)(8.71005,1.96247)
\path(8.89456,2.01917)(8.71005,1.94743)
\path(8.92092,2.01438)(8.78199,1.96036)
\path(8.94728,2.00959)(8.85466,1.97357)
\path(8.97364,2.00479)(8.92733,1.98679)
\path(8.71005,2)(8.71005,1.94728)(9,2)(8.71005,2.05272)(8.71005,2)
\put(1,3){\color{rgb_000000}$\allinethickness{0.070292cm}\circle{0.070292}$}%
\put(1,4){\color{rgb_000000}$\allinethickness{0.070292cm}\circle{0.070292}$}%
\put(2,3){\color{rgb_000000}$\allinethickness{0.070292cm}\circle{0.070292}$}%
\put(2,4){\color{rgb_000000}$\allinethickness{0.070292cm}\circle{0.070292}$}%
\put(3,3){\color{rgb_000000}$\allinethickness{0.070292cm}\circle{0.070292}$}%
\put(3,4){\color{rgb_000000}$\allinethickness{0.070292cm}\circle{0.070292}$}%
\put(4,3){\color{rgb_000000}$\allinethickness{0.070292cm}\circle{0.070292}$}%
\put(4,4){\color{rgb_000000}$\allinethickness{0.070292cm}\circle{0.070292}$}%
\put(5,3){\color{rgb_000000}$\allinethickness{0.070292cm}\circle{0.070292}$}%
\put(5,4){\color{rgb_000000}$\allinethickness{0.070292cm}\circle{0.070292}$}%
\put(6,3){\makebox(0,0)[c]{\hbox{\color{rgb_000000}\scriptsize $\cdots$}}}
\put(6,4){\makebox(0,0)[c]{\hbox{\color{rgb_000000}\scriptsize $\cdots$}}}
\put(7,3){\color{rgb_000000}$\allinethickness{0.070292cm}\circle{0.070292}$}%
\put(7,4){\color{rgb_000000}$\allinethickness{0.070292cm}\circle{0.070292}$}%
\put(8,3){\color{rgb_000000}$\allinethickness{0.070292cm}\circle{0.070292}$}%
\put(8,4){\color{rgb_000000}$\allinethickness{0.070292cm}\circle{0.070292}$}%
\put(2,0.1){\color{rgb_000000}$\allinethickness{0.070292cm}\circle{0.070292}$}%
\put(4,0.9){\color{rgb_000000}$\allinethickness{0.070292cm}\circle{0.070292}$}%
\put(7,0.5){\color{rgb_000000}$\allinethickness{0.070292cm}\circle{0.070292}$}%
\put(1.5,3){\makebox(0,0)[b]{\hbox{\color{rgb_000000}\scriptsize $\longrightarrow$}}}
\put(1.5,3){\makebox(0,0)[t]{\hbox{\color{rgb_000000}\scriptsize $\longleftarrow$}}}
\put(1.5,4){\makebox(0,0)[b]{\hbox{\color{rgb_000000}\scriptsize $\longrightarrow$}}}
\put(1.5,4){\makebox(0,0)[t]{\hbox{\color{rgb_000000}\scriptsize $\longleftarrow$}}}
\put(2.5,3){\makebox(0,0)[b]{\hbox{\color{rgb_000000}\scriptsize $\longrightarrow$}}}
\put(2.5,3){\makebox(0,0)[t]{\hbox{\color{rgb_000000}\scriptsize $\longleftarrow$}}}
\put(2.5,4){\makebox(0,0)[b]{\hbox{\color{rgb_000000}\scriptsize $\longrightarrow$}}}
\put(2.5,4){\makebox(0,0)[t]{\hbox{\color{rgb_000000}\scriptsize $\longleftarrow$}}}
\put(3.5,3){\makebox(0,0)[b]{\hbox{\color{rgb_000000}\scriptsize $\longrightarrow$}}}
\put(3.5,3){\makebox(0,0)[t]{\hbox{\color{rgb_000000}\scriptsize $\longleftarrow$}}}
\put(3.5,4){\makebox(0,0)[b]{\hbox{\color{rgb_000000}\scriptsize $\longrightarrow$}}}
\put(3.5,4){\makebox(0,0)[t]{\hbox{\color{rgb_000000}\scriptsize $\longleftarrow$}}}
\put(4.5,3){\makebox(0,0)[b]{\hbox{\color{rgb_000000}\scriptsize $\longrightarrow$}}}
\put(4.5,3){\makebox(0,0)[t]{\hbox{\color{rgb_000000}\scriptsize $\longleftarrow$}}}
\put(4.5,4){\makebox(0,0)[b]{\hbox{\color{rgb_000000}\scriptsize $\longrightarrow$}}}
\put(4.5,4){\makebox(0,0)[t]{\hbox{\color{rgb_000000}\scriptsize $\longleftarrow$}}}
\put(5.5,3){\makebox(0,0)[b]{\hbox{\color{rgb_000000}\scriptsize $\longrightarrow$}}}
\put(5.5,3){\makebox(0,0)[t]{\hbox{\color{rgb_000000}\scriptsize $\longleftarrow$}}}
\put(5.5,4){\makebox(0,0)[b]{\hbox{\color{rgb_000000}\scriptsize $\longrightarrow$}}}
\put(5.5,4){\makebox(0,0)[t]{\hbox{\color{rgb_000000}\scriptsize $\longleftarrow$}}}
\put(6.5,3){\makebox(0,0)[b]{\hbox{\color{rgb_000000}\scriptsize $\longrightarrow$}}}
\put(6.5,3){\makebox(0,0)[t]{\hbox{\color{rgb_000000}\scriptsize $\longleftarrow$}}}
\put(6.5,4){\makebox(0,0)[b]{\hbox{\color{rgb_000000}\scriptsize $\longrightarrow$}}}
\put(6.5,4){\makebox(0,0)[t]{\hbox{\color{rgb_000000}\scriptsize $\longleftarrow$}}}
\put(7.5,3){\makebox(0,0)[b]{\hbox{\color{rgb_000000}\scriptsize $\longrightarrow$}}}
\put(7.5,3){\makebox(0,0)[t]{\hbox{\color{rgb_000000}\scriptsize $\longleftarrow$}}}
\put(7.5,4){\makebox(0,0)[b]{\hbox{\color{rgb_000000}\scriptsize $\longrightarrow$}}}
\put(7.5,4){\makebox(0,0)[t]{\hbox{\color{rgb_000000}\scriptsize $\longleftarrow$}}}
\put(0,5.07029){\makebox(0,0)[b]{\hbox{\color{rgb_000000}\scriptsize $E$}}}
\put(9.07029,2){\makebox(0,0)[l]{\hbox{\color{rgb_000000}\scriptsize $N_{\text{PF}}$}}}
\put(-0.070292,2){\makebox(0,0)[r]{\hbox{\color{rgb_000000}\scriptsize $E=0$}}}
\put(-0.070292,3){\makebox(0,0)[r]{\hbox{\color{rgb_000000}\scriptsize $E_{1}$}}}
\put(-0.070292,4){\makebox(0,0)[r]{\hbox{\color{rgb_000000}\scriptsize $E_{2}$}}}
\put(-0.246022,4.7){\makebox(0,0)[r]{\hbox{\color{rgb_000000}\scriptsize $\vdots$}}}
\put(1,1.92971){\makebox(0,0)[t]{\hbox{\color{rgb_000000}\scriptsize $0$}}}
\put(2,1.92971){\makebox(0,0)[t]{\hbox{\color{rgb_000000}\scriptsize $1$}}}
\put(3,1.92971){\makebox(0,0)[t]{\hbox{\color{rgb_000000}\scriptsize $2$}}}
\put(4,1.92971){\makebox(0,0)[t]{\hbox{\color{rgb_000000}\scriptsize $3$}}}
\put(5,1.92971){\makebox(0,0)[t]{\hbox{\color{rgb_000000}\scriptsize $4$}}}
\put(7,1.92971){\makebox(0,0)[t]{\hbox{\color{rgb_000000}\scriptsize $N-2$}}}
\put(8,1.92971){\makebox(0,0)[t]{\hbox{\color{rgb_000000}\scriptsize $N-1$}}}
\put(1,3.10544){\makebox(0,0)[b]{\hbox{\color{rgb_000000}\scriptsize $\Vec{\psi}_{0,1}$}}}
\put(2,3.10544){\makebox(0,0)[b]{\hbox{\color{rgb_000000}\scriptsize $\Vec{\psi}_{1,1}$}}}
\put(3,3.10544){\makebox(0,0)[b]{\hbox{\color{rgb_000000}\scriptsize $\Vec{\psi}_{2,1}$}}}
\put(4,3.10544){\makebox(0,0)[b]{\hbox{\color{rgb_000000}\scriptsize $\Vec{\psi}_{3,1}$}}}
\put(5,3.10544){\makebox(0,0)[b]{\hbox{\color{rgb_000000}\scriptsize $\Vec{\psi}_{4,1}$}}}
\put(7,3.10544){\makebox(0,0)[b]{\hbox{\color{rgb_000000}\scriptsize $\Vec{\psi}_{N-2,1}$}}}
\put(8,3.10544){\makebox(0,0)[b]{\hbox{\color{rgb_000000}\scriptsize $\Vec{\psi}_{N-1,1}$}}}
\put(1,4.10544){\makebox(0,0)[b]{\hbox{\color{rgb_000000}\scriptsize $\Vec{\psi}_{0,2}$}}}
\put(2,4.10544){\makebox(0,0)[b]{\hbox{\color{rgb_000000}\scriptsize $\Vec{\psi}_{1,2}$}}}
\put(3,4.10544){\makebox(0,0)[b]{\hbox{\color{rgb_000000}\scriptsize $\Vec{\psi}_{2,2}$}}}
\put(4,4.10544){\makebox(0,0)[b]{\hbox{\color{rgb_000000}\scriptsize $\Vec{\psi}_{3,2}$}}}
\put(5,4.10544){\makebox(0,0)[b]{\hbox{\color{rgb_000000}\scriptsize $\Vec{\psi}_{4,2}$}}}
\put(7,4.10544){\makebox(0,0)[b]{\hbox{\color{rgb_000000}\scriptsize $\Vec{\psi}_{N-2,2}$}}}
\put(8,4.10544){\makebox(0,0)[b]{\hbox{\color{rgb_000000}\scriptsize $\Vec{\psi}_{N-1,2}$}}}
\put(2,0.205438){\makebox(0,0)[b]{\hbox{\color{rgb_000000}\scriptsize $\Vec{\psi}_{1,0}$}}}
\put(4,1.00544){\makebox(0,0)[b]{\hbox{\color{rgb_000000}\scriptsize $\Vec{\psi}_{3,0}$}}}
\put(7,0.605438){\makebox(0,0)[b]{\hbox{\color{rgb_000000}\scriptsize $\Vec{\psi}_{N-2,0}$}}}
\put(1.82427,0.1){\makebox(0,0)[r]{\hbox{\color{rgb_000000}\scriptsize $E_{1,0}$}}}
\put(3.82427,0.9){\makebox(0,0)[r]{\hbox{\color{rgb_000000}\scriptsize $E_{3,0}$}}}
\put(6.82427,0.5){\makebox(0,0)[r]{\hbox{\color{rgb_000000}\scriptsize $E_{N-2,0}$}}}
\end{picture}%
 \label{fig:2b}}
\caption{Schematic parasuperspectrum for the type (D, D) boundary conditions. Parasupersymmetry is unbroken.}
\label{fig:2}
\end{figure}

\begin{figure}[t]
\centering
\subfigure[Case $N$ even.]{
\xdefinecolor{rgb_000000}{rgb}{0,0,0}%
\setlength{\unitlength}{1cm}%
\begin{picture}(8,5)(0,0)%
\path(0,0)(0,5)
\path(-0.052719,4.71005)(0.052719,4.71005)
\path(-0.0502085,4.72385)(0.0502085,4.72385)
\path(-0.0476981,4.73766)(0.0476981,4.73766)
\path(-0.0451877,4.75147)(0.0451877,4.75147)
\path(-0.0426773,4.76528)(0.0426773,4.76528)
\path(-0.0401668,4.77908)(0.0401668,4.77908)
\path(-0.0376564,4.79289)(0.0376564,4.79289)
\path(-0.035146,4.8067)(0.035146,4.8067)
\path(-0.0326356,4.8205)(0.0326356,4.8205)
\path(-0.0301251,4.83431)(0.0301251,4.83431)
\path(-0.0276147,4.84812)(0.0276147,4.84812)
\path(-0.0251043,4.86193)(0.0251043,4.86193)
\path(-0.0225938,4.87573)(0.0225938,4.87573)
\path(-0.0200834,4.88954)(0.0200834,4.88954)
\path(-0.017573,4.90335)(0.017573,4.90335)
\path(-0.0150626,4.91716)(0.0150626,4.91716)
\path(-0.0125521,4.93096)(0.0125521,4.93096)
\path(-0.0100417,4.94477)(0.0100417,4.94477)
\path(-0.00753128,4.95858)(0.00753128,4.95858)
\path(-0.00502085,4.97239)(0.00502085,4.97239)
\path(-0.00251043,4.98619)(0.00251043,4.98619)
\path(0.0395392,4.71005)(0.0395392,4.78253)
\path(0.0263595,4.71005)(0.0263595,4.85502)
\path(0.0131797,4.71005)(0.0131797,4.92751)
\path(0,4.71005)(0,5)
\path(-0.0131797,4.71005)(-0.0131797,4.92751)
\path(-0.0263595,4.71005)(-0.0263595,4.85502)
\path(-0.0395392,4.71005)(-0.0395392,4.78253)
\path(0,4.71005)(0.052719,4.71005)(0,5)(-0.052719,4.71005)(0,4.71005)
\path(0,2)(8,2)
\path(7.98619,1.99749)(7.98409,2.00289)
\path(7.97239,1.99498)(7.96818,2.00578)
\path(7.95858,1.99247)(7.95228,2.00868)
\path(7.94477,1.98996)(7.93637,2.01157)
\path(7.93096,1.98745)(7.92046,2.01446)
\path(7.91716,1.98494)(7.90455,2.01735)
\path(7.90335,1.98243)(7.88864,2.02025)
\path(7.88954,1.97992)(7.87274,2.02314)
\path(7.87573,1.97741)(7.85683,2.02603)
\path(7.86193,1.9749)(7.84092,2.02892)
\path(7.84812,1.97239)(7.82501,2.03182)
\path(7.83431,1.96987)(7.8091,2.03471)
\path(7.8205,1.96736)(7.7932,2.0376)
\path(7.8067,1.96485)(7.77729,2.04049)
\path(7.79289,1.96234)(7.76138,2.04339)
\path(7.77908,1.95983)(7.74547,2.04628)
\path(7.76528,1.95732)(7.72956,2.04917)
\path(7.75147,1.95481)(7.71366,2.05206)
\path(7.73766,1.9523)(7.71005,2.02333)
\path(7.72385,1.94979)(7.71005,1.9853)
\path(7.73641,2.04793)(7.71005,2.03768)
\path(7.76276,2.04313)(7.71005,2.02264)
\path(7.78912,2.03834)(7.71005,2.0076)
\path(7.81548,2.03355)(7.71005,1.99255)
\path(7.84184,2.02876)(7.71005,1.97751)
\path(7.8682,2.02396)(7.71005,1.96247)
\path(7.89456,2.01917)(7.71005,1.94743)
\path(7.92092,2.01438)(7.78199,1.96036)
\path(7.94728,2.00959)(7.85466,1.97357)
\path(7.97364,2.00479)(7.92733,1.98679)
\path(7.71005,2)(7.71005,1.94728)(8,2)(7.71005,2.05272)(7.71005,2)
\put(1,3){\color{rgb_000000}$\allinethickness{0.070292cm}\circle{0.070292}$}%
\put(1,4){\color{rgb_000000}$\allinethickness{0.070292cm}\circle{0.070292}$}%
\put(2,3){\color{rgb_000000}$\allinethickness{0.070292cm}\circle{0.070292}$}%
\put(2,4){\color{rgb_000000}$\allinethickness{0.070292cm}\circle{0.070292}$}%
\put(3,3){\color{rgb_000000}$\allinethickness{0.070292cm}\circle{0.070292}$}%
\put(3,4){\color{rgb_000000}$\allinethickness{0.070292cm}\circle{0.070292}$}%
\put(4,3){\color{rgb_000000}$\allinethickness{0.070292cm}\circle{0.070292}$}%
\put(4,4){\color{rgb_000000}$\allinethickness{0.070292cm}\circle{0.070292}$}%
\put(5,3){\makebox(0,0)[c]{\hbox{\color{rgb_000000}\scriptsize $\cdots$}}}
\put(5,4){\makebox(0,0)[c]{\hbox{\color{rgb_000000}\scriptsize $\cdots$}}}
\put(6,3){\color{rgb_000000}$\allinethickness{0.070292cm}\circle{0.070292}$}%
\put(6,4){\color{rgb_000000}$\allinethickness{0.070292cm}\circle{0.070292}$}%
\put(7,3){\color{rgb_000000}$\allinethickness{0.070292cm}\circle{0.070292}$}%
\put(7,4){\color{rgb_000000}$\allinethickness{0.070292cm}\circle{0.070292}$}%
\put(1,0.5){\color{rgb_000000}$\allinethickness{0.070292cm}\circle{0.070292}$}%
\put(3,0.1){\color{rgb_000000}$\allinethickness{0.070292cm}\circle{0.070292}$}%
\put(6,0.9){\color{rgb_000000}$\allinethickness{0.070292cm}\circle{0.070292}$}%
\put(1.5,3){\makebox(0,0)[b]{\hbox{\color{rgb_000000}\scriptsize $\longrightarrow$}}}
\put(1.5,3){\makebox(0,0)[t]{\hbox{\color{rgb_000000}\scriptsize $\longleftarrow$}}}
\put(1.5,4){\makebox(0,0)[b]{\hbox{\color{rgb_000000}\scriptsize $\longrightarrow$}}}
\put(1.5,4){\makebox(0,0)[t]{\hbox{\color{rgb_000000}\scriptsize $\longleftarrow$}}}
\put(2.5,3){\makebox(0,0)[b]{\hbox{\color{rgb_000000}\scriptsize $\longrightarrow$}}}
\put(2.5,3){\makebox(0,0)[t]{\hbox{\color{rgb_000000}\scriptsize $\longleftarrow$}}}
\put(2.5,4){\makebox(0,0)[b]{\hbox{\color{rgb_000000}\scriptsize $\longrightarrow$}}}
\put(2.5,4){\makebox(0,0)[t]{\hbox{\color{rgb_000000}\scriptsize $\longleftarrow$}}}
\put(3.5,3){\makebox(0,0)[b]{\hbox{\color{rgb_000000}\scriptsize $\longrightarrow$}}}
\put(3.5,3){\makebox(0,0)[t]{\hbox{\color{rgb_000000}\scriptsize $\longleftarrow$}}}
\put(3.5,4){\makebox(0,0)[b]{\hbox{\color{rgb_000000}\scriptsize $\longrightarrow$}}}
\put(3.5,4){\makebox(0,0)[t]{\hbox{\color{rgb_000000}\scriptsize $\longleftarrow$}}}
\put(4.5,3){\makebox(0,0)[b]{\hbox{\color{rgb_000000}\scriptsize $\longrightarrow$}}}
\put(4.5,3){\makebox(0,0)[t]{\hbox{\color{rgb_000000}\scriptsize $\longleftarrow$}}}
\put(4.5,4){\makebox(0,0)[b]{\hbox{\color{rgb_000000}\scriptsize $\longrightarrow$}}}
\put(4.5,4){\makebox(0,0)[t]{\hbox{\color{rgb_000000}\scriptsize $\longleftarrow$}}}
\put(5.5,3){\makebox(0,0)[b]{\hbox{\color{rgb_000000}\scriptsize $\longrightarrow$}}}
\put(5.5,3){\makebox(0,0)[t]{\hbox{\color{rgb_000000}\scriptsize $\longleftarrow$}}}
\put(5.5,4){\makebox(0,0)[b]{\hbox{\color{rgb_000000}\scriptsize $\longrightarrow$}}}
\put(5.5,4){\makebox(0,0)[t]{\hbox{\color{rgb_000000}\scriptsize $\longleftarrow$}}}
\put(6.5,3){\makebox(0,0)[b]{\hbox{\color{rgb_000000}\scriptsize $\longrightarrow$}}}
\put(6.5,3){\makebox(0,0)[t]{\hbox{\color{rgb_000000}\scriptsize $\longleftarrow$}}}
\put(6.5,4){\makebox(0,0)[b]{\hbox{\color{rgb_000000}\scriptsize $\longrightarrow$}}}
\put(6.5,4){\makebox(0,0)[t]{\hbox{\color{rgb_000000}\scriptsize $\longleftarrow$}}}
\put(0,5.07029){\makebox(0,0)[b]{\hbox{\color{rgb_000000}\scriptsize $E$}}}
\put(8.07029,2){\makebox(0,0)[l]{\hbox{\color{rgb_000000}\scriptsize $N_{\text{PF}}$}}}
\put(-0.070292,2){\makebox(0,0)[r]{\hbox{\color{rgb_000000}\scriptsize $E=0$}}}
\put(-0.070292,3){\makebox(0,0)[r]{\hbox{\color{rgb_000000}\scriptsize $E_{1}$}}}
\put(-0.070292,4){\makebox(0,0)[r]{\hbox{\color{rgb_000000}\scriptsize $E_{2}$}}}
\put(-0.246022,4.7){\makebox(0,0)[r]{\hbox{\color{rgb_000000}\scriptsize $\vdots$}}}
\put(1,1.92971){\makebox(0,0)[t]{\hbox{\color{rgb_000000}\scriptsize $0$}}}
\put(2,1.92971){\makebox(0,0)[t]{\hbox{\color{rgb_000000}\scriptsize $1$}}}
\put(3,1.92971){\makebox(0,0)[t]{\hbox{\color{rgb_000000}\scriptsize $2$}}}
\put(4,1.92971){\makebox(0,0)[t]{\hbox{\color{rgb_000000}\scriptsize $3$}}}
\put(6,1.92971){\makebox(0,0)[t]{\hbox{\color{rgb_000000}\scriptsize $N-2$}}}
\put(7,1.92971){\makebox(0,0)[t]{\hbox{\color{rgb_000000}\scriptsize $N-1$}}}
\put(1,3.10544){\makebox(0,0)[b]{\hbox{\color{rgb_000000}\scriptsize $\Vec{\psi}_{0,1}$}}}
\put(2,3.10544){\makebox(0,0)[b]{\hbox{\color{rgb_000000}\scriptsize $\Vec{\psi}_{1,1}$}}}
\put(3,3.10544){\makebox(0,0)[b]{\hbox{\color{rgb_000000}\scriptsize $\Vec{\psi}_{2,1}$}}}
\put(4,3.10544){\makebox(0,0)[b]{\hbox{\color{rgb_000000}\scriptsize $\Vec{\psi}_{3,1}$}}}
\put(6,3.10544){\makebox(0,0)[b]{\hbox{\color{rgb_000000}\scriptsize $\Vec{\psi}_{N-2,1}$}}}
\put(7,3.10544){\makebox(0,0)[b]{\hbox{\color{rgb_000000}\scriptsize $\Vec{\psi}_{N-1,1}$}}}
\put(1,4.10544){\makebox(0,0)[b]{\hbox{\color{rgb_000000}\scriptsize $\Vec{\psi}_{0,2}$}}}
\put(2,4.10544){\makebox(0,0)[b]{\hbox{\color{rgb_000000}\scriptsize $\Vec{\psi}_{1,2}$}}}
\put(3,4.10544){\makebox(0,0)[b]{\hbox{\color{rgb_000000}\scriptsize $\Vec{\psi}_{2,2}$}}}
\put(4,4.10544){\makebox(0,0)[b]{\hbox{\color{rgb_000000}\scriptsize $\Vec{\psi}_{3,2}$}}}
\put(6,4.10544){\makebox(0,0)[b]{\hbox{\color{rgb_000000}\scriptsize $\Vec{\psi}_{N-2,2}$}}}
\put(7,4.10544){\makebox(0,0)[b]{\hbox{\color{rgb_000000}\scriptsize $\Vec{\psi}_{N-1,2}$}}}
\put(1,0.605438){\makebox(0,0)[b]{\hbox{\color{rgb_000000}\scriptsize $\Vec{\psi}_{0,0}$}}}
\put(3,0.205438){\makebox(0,0)[b]{\hbox{\color{rgb_000000}\scriptsize $\Vec{\psi}_{2,0}$}}}
\put(6,1.00544){\makebox(0,0)[b]{\hbox{\color{rgb_000000}\scriptsize $\Vec{\psi}_{N-2,0}$}}}
\put(0.82427,0.5){\makebox(0,0)[r]{\hbox{\color{rgb_000000}\scriptsize $E_{0,0}$}}}
\put(2.82427,0.1){\makebox(0,0)[r]{\hbox{\color{rgb_000000}\scriptsize $E_{2,0}$}}}
\put(5.82427,0.9){\makebox(0,0)[r]{\hbox{\color{rgb_000000}\scriptsize $E_{N-2,0}$}}}
\end{picture}%
 \label{fig:3a}} \hspace{1cm}
\subfigure[Case $N$ odd.]{
\xdefinecolor{rgb_000000}{rgb}{0,0,0}%
\setlength{\unitlength}{1cm}%
\begin{picture}(9,5)(0,0)%
\path(0,0)(0,5)
\path(-0.052719,4.71005)(0.052719,4.71005)
\path(-0.0502085,4.72385)(0.0502085,4.72385)
\path(-0.0476981,4.73766)(0.0476981,4.73766)
\path(-0.0451877,4.75147)(0.0451877,4.75147)
\path(-0.0426773,4.76528)(0.0426773,4.76528)
\path(-0.0401668,4.77908)(0.0401668,4.77908)
\path(-0.0376564,4.79289)(0.0376564,4.79289)
\path(-0.035146,4.8067)(0.035146,4.8067)
\path(-0.0326356,4.8205)(0.0326356,4.8205)
\path(-0.0301251,4.83431)(0.0301251,4.83431)
\path(-0.0276147,4.84812)(0.0276147,4.84812)
\path(-0.0251043,4.86193)(0.0251043,4.86193)
\path(-0.0225938,4.87573)(0.0225938,4.87573)
\path(-0.0200834,4.88954)(0.0200834,4.88954)
\path(-0.017573,4.90335)(0.017573,4.90335)
\path(-0.0150626,4.91716)(0.0150626,4.91716)
\path(-0.0125521,4.93096)(0.0125521,4.93096)
\path(-0.0100417,4.94477)(0.0100417,4.94477)
\path(-0.00753128,4.95858)(0.00753128,4.95858)
\path(-0.00502085,4.97239)(0.00502085,4.97239)
\path(-0.00251043,4.98619)(0.00251043,4.98619)
\path(0.0395392,4.71005)(0.0395392,4.78253)
\path(0.0263595,4.71005)(0.0263595,4.85502)
\path(0.0131797,4.71005)(0.0131797,4.92751)
\path(0,4.71005)(0,5)
\path(-0.0131797,4.71005)(-0.0131797,4.92751)
\path(-0.0263595,4.71005)(-0.0263595,4.85502)
\path(-0.0395392,4.71005)(-0.0395392,4.78253)
\path(0,4.71005)(0.052719,4.71005)(0,5)(-0.052719,4.71005)(0,4.71005)
\path(0,2)(9,2)
\path(8.98619,1.99749)(8.98409,2.00289)
\path(8.97239,1.99498)(8.96818,2.00578)
\path(8.95858,1.99247)(8.95228,2.00868)
\path(8.94477,1.98996)(8.93637,2.01157)
\path(8.93096,1.98745)(8.92046,2.01446)
\path(8.91716,1.98494)(8.90455,2.01735)
\path(8.90335,1.98243)(8.88864,2.02025)
\path(8.88954,1.97992)(8.87274,2.02314)
\path(8.87573,1.97741)(8.85683,2.02603)
\path(8.86193,1.9749)(8.84092,2.02892)
\path(8.84812,1.97239)(8.82501,2.03182)
\path(8.83431,1.96987)(8.8091,2.03471)
\path(8.8205,1.96736)(8.7932,2.0376)
\path(8.8067,1.96485)(8.77729,2.04049)
\path(8.79289,1.96234)(8.76138,2.04339)
\path(8.77908,1.95983)(8.74547,2.04628)
\path(8.76528,1.95732)(8.72956,2.04917)
\path(8.75147,1.95481)(8.71366,2.05206)
\path(8.73766,1.9523)(8.71005,2.02333)
\path(8.72385,1.94979)(8.71005,1.9853)
\path(8.73641,2.04793)(8.71005,2.03768)
\path(8.76276,2.04313)(8.71005,2.02264)
\path(8.78912,2.03834)(8.71005,2.0076)
\path(8.81548,2.03355)(8.71005,1.99255)
\path(8.84184,2.02876)(8.71005,1.97751)
\path(8.8682,2.02396)(8.71005,1.96247)
\path(8.89456,2.01917)(8.71005,1.94743)
\path(8.92092,2.01438)(8.78199,1.96036)
\path(8.94728,2.00959)(8.85466,1.97357)
\path(8.97364,2.00479)(8.92733,1.98679)
\path(8.71005,2)(8.71005,1.94728)(9,2)(8.71005,2.05272)(8.71005,2)
\put(1,3){\color{rgb_000000}$\allinethickness{0.070292cm}\circle{0.070292}$}%
\put(1,4){\color{rgb_000000}$\allinethickness{0.070292cm}\circle{0.070292}$}%
\put(2,3){\color{rgb_000000}$\allinethickness{0.070292cm}\circle{0.070292}$}%
\put(2,4){\color{rgb_000000}$\allinethickness{0.070292cm}\circle{0.070292}$}%
\put(3,3){\color{rgb_000000}$\allinethickness{0.070292cm}\circle{0.070292}$}%
\put(3,4){\color{rgb_000000}$\allinethickness{0.070292cm}\circle{0.070292}$}%
\put(4,3){\color{rgb_000000}$\allinethickness{0.070292cm}\circle{0.070292}$}%
\put(4,4){\color{rgb_000000}$\allinethickness{0.070292cm}\circle{0.070292}$}%
\put(5,3){\color{rgb_000000}$\allinethickness{0.070292cm}\circle{0.070292}$}%
\put(5,4){\color{rgb_000000}$\allinethickness{0.070292cm}\circle{0.070292}$}%
\put(6,3){\makebox(0,0)[c]{\hbox{\color{rgb_000000}\scriptsize $\cdots$}}}
\put(6,4){\makebox(0,0)[c]{\hbox{\color{rgb_000000}\scriptsize $\cdots$}}}
\put(7,3){\color{rgb_000000}$\allinethickness{0.070292cm}\circle{0.070292}$}%
\put(7,4){\color{rgb_000000}$\allinethickness{0.070292cm}\circle{0.070292}$}%
\put(8,3){\color{rgb_000000}$\allinethickness{0.070292cm}\circle{0.070292}$}%
\put(8,4){\color{rgb_000000}$\allinethickness{0.070292cm}\circle{0.070292}$}%
\put(1,0.9){\color{rgb_000000}$\allinethickness{0.070292cm}\circle{0.070292}$}%
\put(3,1.3){\color{rgb_000000}$\allinethickness{0.070292cm}\circle{0.070292}$}%
\put(5,0.1){\color{rgb_000000}$\allinethickness{0.070292cm}\circle{0.070292}$}%
\put(8,0.5){\color{rgb_000000}$\allinethickness{0.070292cm}\circle{0.070292}$}%
\put(1.5,3){\makebox(0,0)[b]{\hbox{\color{rgb_000000}\scriptsize $\longrightarrow$}}}
\put(1.5,3){\makebox(0,0)[t]{\hbox{\color{rgb_000000}\scriptsize $\longleftarrow$}}}
\put(1.5,4){\makebox(0,0)[b]{\hbox{\color{rgb_000000}\scriptsize $\longrightarrow$}}}
\put(1.5,4){\makebox(0,0)[t]{\hbox{\color{rgb_000000}\scriptsize $\longleftarrow$}}}
\put(2.5,3){\makebox(0,0)[b]{\hbox{\color{rgb_000000}\scriptsize $\longrightarrow$}}}
\put(2.5,3){\makebox(0,0)[t]{\hbox{\color{rgb_000000}\scriptsize $\longleftarrow$}}}
\put(2.5,4){\makebox(0,0)[b]{\hbox{\color{rgb_000000}\scriptsize $\longrightarrow$}}}
\put(2.5,4){\makebox(0,0)[t]{\hbox{\color{rgb_000000}\scriptsize $\longleftarrow$}}}
\put(3.5,3){\makebox(0,0)[b]{\hbox{\color{rgb_000000}\scriptsize $\longrightarrow$}}}
\put(3.5,3){\makebox(0,0)[t]{\hbox{\color{rgb_000000}\scriptsize $\longleftarrow$}}}
\put(3.5,4){\makebox(0,0)[b]{\hbox{\color{rgb_000000}\scriptsize $\longrightarrow$}}}
\put(3.5,4){\makebox(0,0)[t]{\hbox{\color{rgb_000000}\scriptsize $\longleftarrow$}}}
\put(4.5,3){\makebox(0,0)[b]{\hbox{\color{rgb_000000}\scriptsize $\longrightarrow$}}}
\put(4.5,3){\makebox(0,0)[t]{\hbox{\color{rgb_000000}\scriptsize $\longleftarrow$}}}
\put(4.5,4){\makebox(0,0)[b]{\hbox{\color{rgb_000000}\scriptsize $\longrightarrow$}}}
\put(4.5,4){\makebox(0,0)[t]{\hbox{\color{rgb_000000}\scriptsize $\longleftarrow$}}}
\put(5.5,3){\makebox(0,0)[b]{\hbox{\color{rgb_000000}\scriptsize $\longrightarrow$}}}
\put(5.5,3){\makebox(0,0)[t]{\hbox{\color{rgb_000000}\scriptsize $\longleftarrow$}}}
\put(5.5,4){\makebox(0,0)[b]{\hbox{\color{rgb_000000}\scriptsize $\longrightarrow$}}}
\put(5.5,4){\makebox(0,0)[t]{\hbox{\color{rgb_000000}\scriptsize $\longleftarrow$}}}
\put(6.5,3){\makebox(0,0)[b]{\hbox{\color{rgb_000000}\scriptsize $\longrightarrow$}}}
\put(6.5,3){\makebox(0,0)[t]{\hbox{\color{rgb_000000}\scriptsize $\longleftarrow$}}}
\put(6.5,4){\makebox(0,0)[b]{\hbox{\color{rgb_000000}\scriptsize $\longrightarrow$}}}
\put(6.5,4){\makebox(0,0)[t]{\hbox{\color{rgb_000000}\scriptsize $\longleftarrow$}}}
\put(7.5,3){\makebox(0,0)[b]{\hbox{\color{rgb_000000}\scriptsize $\longrightarrow$}}}
\put(7.5,3){\makebox(0,0)[t]{\hbox{\color{rgb_000000}\scriptsize $\longleftarrow$}}}
\put(7.5,4){\makebox(0,0)[b]{\hbox{\color{rgb_000000}\scriptsize $\longrightarrow$}}}
\put(7.5,4){\makebox(0,0)[t]{\hbox{\color{rgb_000000}\scriptsize $\longleftarrow$}}}
\put(0,5.07029){\makebox(0,0)[b]{\hbox{\color{rgb_000000}\scriptsize $E$}}}
\put(9.07029,2){\makebox(0,0)[l]{\hbox{\color{rgb_000000}\scriptsize $N_{\text{PF}}$}}}
\put(-0.070292,2){\makebox(0,0)[r]{\hbox{\color{rgb_000000}\scriptsize $E=0$}}}
\put(-0.070292,3){\makebox(0,0)[r]{\hbox{\color{rgb_000000}\scriptsize $E_{1}$}}}
\put(-0.070292,4){\makebox(0,0)[r]{\hbox{\color{rgb_000000}\scriptsize $E_{2}$}}}
\put(-0.246022,4.7){\makebox(0,0)[r]{\hbox{\color{rgb_000000}\scriptsize $\vdots$}}}
\put(1,1.92971){\makebox(0,0)[t]{\hbox{\color{rgb_000000}\scriptsize $0$}}}
\put(2,1.92971){\makebox(0,0)[t]{\hbox{\color{rgb_000000}\scriptsize $1$}}}
\put(3,1.92971){\makebox(0,0)[t]{\hbox{\color{rgb_000000}\scriptsize $2$}}}
\put(4,1.92971){\makebox(0,0)[t]{\hbox{\color{rgb_000000}\scriptsize $3$}}}
\put(5,1.92971){\makebox(0,0)[t]{\hbox{\color{rgb_000000}\scriptsize $4$}}}
\put(7,1.92971){\makebox(0,0)[t]{\hbox{\color{rgb_000000}\scriptsize $N-2$}}}
\put(8,1.92971){\makebox(0,0)[t]{\hbox{\color{rgb_000000}\scriptsize $N-1$}}}
\put(1,3.10544){\makebox(0,0)[b]{\hbox{\color{rgb_000000}\scriptsize $\Vec{\psi}_{0,1}$}}}
\put(2,3.10544){\makebox(0,0)[b]{\hbox{\color{rgb_000000}\scriptsize $\Vec{\psi}_{1,1}$}}}
\put(3,3.10544){\makebox(0,0)[b]{\hbox{\color{rgb_000000}\scriptsize $\Vec{\psi}_{2,1}$}}}
\put(4,3.10544){\makebox(0,0)[b]{\hbox{\color{rgb_000000}\scriptsize $\Vec{\psi}_{3,1}$}}}
\put(5,3.10544){\makebox(0,0)[b]{\hbox{\color{rgb_000000}\scriptsize $\Vec{\psi}_{4,1}$}}}
\put(7,3.10544){\makebox(0,0)[b]{\hbox{\color{rgb_000000}\scriptsize $\Vec{\psi}_{N-2,1}$}}}
\put(8,3.10544){\makebox(0,0)[b]{\hbox{\color{rgb_000000}\scriptsize $\Vec{\psi}_{N-1,1}$}}}
\put(1,4.10544){\makebox(0,0)[b]{\hbox{\color{rgb_000000}\scriptsize $\Vec{\psi}_{0,2}$}}}
\put(2,4.10544){\makebox(0,0)[b]{\hbox{\color{rgb_000000}\scriptsize $\Vec{\psi}_{1,2}$}}}
\put(3,4.10544){\makebox(0,0)[b]{\hbox{\color{rgb_000000}\scriptsize $\Vec{\psi}_{2,2}$}}}
\put(4,4.10544){\makebox(0,0)[b]{\hbox{\color{rgb_000000}\scriptsize $\Vec{\psi}_{3,2}$}}}
\put(5,4.10544){\makebox(0,0)[b]{\hbox{\color{rgb_000000}\scriptsize $\Vec{\psi}_{4,2}$}}}
\put(7,4.10544){\makebox(0,0)[b]{\hbox{\color{rgb_000000}\scriptsize $\Vec{\psi}_{N-2,2}$}}}
\put(8,4.10544){\makebox(0,0)[b]{\hbox{\color{rgb_000000}\scriptsize $\Vec{\psi}_{N-1,2}$}}}
\put(1,1.00544){\makebox(0,0)[b]{\hbox{\color{rgb_000000}\scriptsize $\Vec{\psi}_{0,0}$}}}
\put(3,1.40544){\makebox(0,0)[b]{\hbox{\color{rgb_000000}\scriptsize $\Vec{\psi}_{2,0}$}}}
\put(5,0.205438){\makebox(0,0)[b]{\hbox{\color{rgb_000000}\scriptsize $\Vec{\psi}_{4,0}$}}}
\put(8,0.605438){\makebox(0,0)[b]{\hbox{\color{rgb_000000}\scriptsize $\Vec{\psi}_{N-1,0}$}}}
\put(0.82427,0.9){\makebox(0,0)[r]{\hbox{\color{rgb_000000}\scriptsize $E_{0,0}$}}}
\put(2.82427,1.3){\makebox(0,0)[r]{\hbox{\color{rgb_000000}\scriptsize $E_{2,0}$}}}
\put(4.82427,0.1){\makebox(0,0)[r]{\hbox{\color{rgb_000000}\scriptsize $E_{4,0}$}}}
\put(7.82427,0.5){\makebox(0,0)[r]{\hbox{\color{rgb_000000}\scriptsize $E_{N-1,0}$}}}
\end{picture}%
 \label{fig:3b}}
\caption{Schematic parasuperspectrum for the type (R, R) boundary conditions. Parasupersymmetry is unbroken.}
\label{fig:3}
\end{figure}
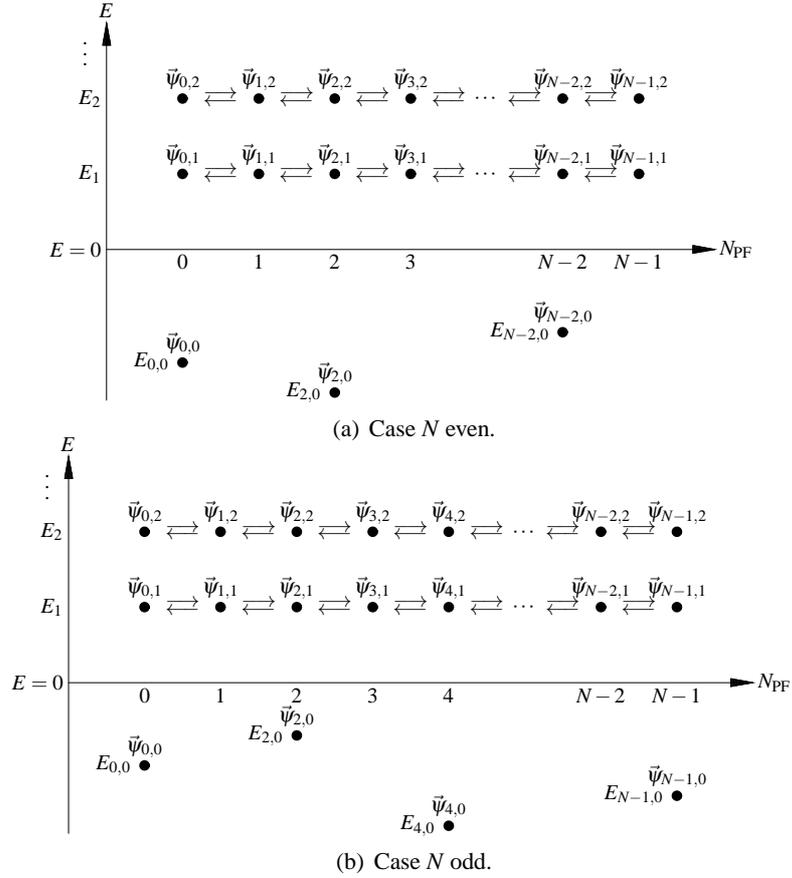

Throughout this section we will concentrate on the case $N$ even.
The case $N$ odd can be similarly analyzed.
Both cases are pictorially summarized in Figures \ref{fig:2}--\ref{fig:4}.

\subsection{Type (D, D) boundary conditions} \label{sec:4.1}
Let us first study the spectrum under the type (D, D) boundary conditions.
To this end, let us first focus on the positive energy eigenstates, which are $N$-fold degenerate thanks to parasupersymmetry.
The normalized energy eigenfunctions are given by
\begin{subequations}
\begin{align}
\Vec{\psi}_{n,\nu}(x) &= (-1)^{\frac{n}{2}}\sqrt{\frac{2}{\ell}}\sin(k_{\nu}x)\Vec{e}_{n} 	& (n=0,2,4,\cdots,N-2), \label{eq:4.1a}\\
\Vec{\psi}_{n,\nu}(x) &= (-1)^{\frac{n-1}{2}}\sqrt{\frac{2/\ell}{k_{\nu}^{2} + M^{2}(\alpha_{n})}}\bigl[M(\alpha_{n})\sin(k_{\nu}x) + k_{\nu}\cos(k_{\nu}x)\bigr]\Vec{e}_{n} 	& (n=1,3,5,\cdots,N-1), \label{eq:4.1b}
\end{align}
\end{subequations}
where $k_{\nu} = \nu\pi/\ell$ ($\nu=1,2,3,\cdots$).
We note that relative phases are fixed to satisfy the parasupersymmetry relations \eqref{eq:3.2a}--\eqref{eq:3.2c}, or, equivalently, $\Vec{\psi}_{n,\nu}(x) = \left[\prod_{m=1}^{n}\bigl(k_{\nu}^{2} + M^{2}(\alpha_{m})\bigr)^{-1/2}\right](Q^{+})^{n}\Vec{\psi}_{0,\nu}(x)$ with the constraints $\alpha_{n} = -\alpha_{n-1}$ ($n=2,4,6,\cdots,N-2$).
The positive energy eigenvalues are given by
\begin{align}
E_{\nu} = \left(\frac{\nu\pi}{\ell}\right)^{2} \quad (\nu=1,2,3,\cdots). \label{eq:4.2}
\end{align}
In addition to these states, there are $N/2$ distinct negative energy eigenstates
\begin{align}
\Vec{\psi}_{n,0}(x) = \sqrt{\frac{2M(\alpha_{n})}{\mathrm{e}^{2M(\alpha_{n})\ell} - 1}}\exp\bigl[M(\alpha_{n})x\bigr]\Vec{e}_{n} \quad (n=1,3,5,\cdots,N-1), \label{eq:4.3}
\end{align}
with the energy eigenvalues
\begin{align}
E_{n,0} = -M^{2}(\alpha_{n}) \quad (n=1,3,5,\cdots,N-1). \label{eq:4.4}
\end{align}
Notice that these negative energy states are non-degenerate in general without tuning the parameters $\{\alpha_{1}, \alpha_{3}, \alpha_{5}, \cdots, \alpha_{N-1}\}$.
Note also that the negative energy eigenstates \eqref{eq:4.3} are the zero-modes of the differential operators $A^{-}_{\alpha_{n}}$ ($n=1,3,5,\cdots,N-1$) and therefore annihilated by the parasupercharges, $Q^{\pm}\Vec{\psi}_{n,0}(x) = \Vec{0}$.
Since the ground state(s) is/are given by the lowest negative energy eigenstate(s), we see that in the type (D, D) boundary conditions parasupersymmetry is unbroken.
Figure \ref{fig:2} schematically shows the spectrum for both $N$ even and odd.

\subsection{Type (R, R) boundary conditions} \label{sec:4.2}
Let us next study the spectrum under the type (R, R) boundary conditions.
The positive energy eigenstates are $N$-fold degenerate and given by
\begin{subequations}
\begin{align}
\Vec{\psi}_{n,\nu}(x) &= (-1)^{\frac{n}{2}}\sqrt{\frac{2/\ell}{k_{\nu}^{2} + M^{2}(\alpha_{n+1})}}\bigl[M(\alpha_{n+1})\sin(k_{\nu}x) - k_{\nu}\cos(k_{\nu}x)\bigr]\Vec{e}_{n} 	& (n=0,2,4,\cdots,N-2), \label{eq:4.5a}\\
\Vec{\psi}_{n,\nu}(x) &= (-1)^{\frac{n-1}{2}}\sqrt{\frac{2}{\ell}}\sin(k_{\nu}x)\Vec{e}_{n} 	& (n=1,3,5,\cdots,N-1), \label{eq:4.5b}
\end{align}
\end{subequations}
where $k_{\nu} = \nu\pi/\ell$ ($\nu=1,2,3,\cdots$).
Again relative phases are fixed by the parasupersymmetry relations with the constraints $\alpha_{n} = -\alpha_{n+1}$ ($n=2,4,6,\cdots,N-2$).
The positive energy eigenvalues are the same as the previous ones
\begin{align}
E_{\nu} = \left(\frac{\nu\pi}{\ell}\right)^{2} \quad (\nu=1,2,3,\cdots). \label{eq:4.6}
\end{align}
Just as in the case of type (D, D) boundary conditions, there are $N/2$ distinct non-degenerate negative energy eigenstates
\begin{align}
\Vec{\psi}_{n,0}(x) = \sqrt{\frac{2M(\alpha_{n+1})}{1 - \mathrm{e}^{-2M(\alpha_{n+1})\ell}}}\exp\bigl[-M(\alpha_{n+1})x\bigr]\Vec{e}_{n} \quad (n=0,2,4,\cdots,N-2), \label{eq:4.7}
\end{align}
whose energy eigenvalues are given by
\begin{align}
E_{n,0} = -M^{2}(\alpha_{n+1}) \quad (n=0,2,4,\cdots,N-2). \label{eq:4.8}
\end{align}
These negative energy states are annihilated by the parasupercharges such that parasupersymmetry is again unbroken.
The resultant spectrum is schematically depicted in Figure \ref{fig:3}.

\begin{figure}[t]
\centering
\subfigure[Case $M(\alpha)\ell \geq -1$.]{\input{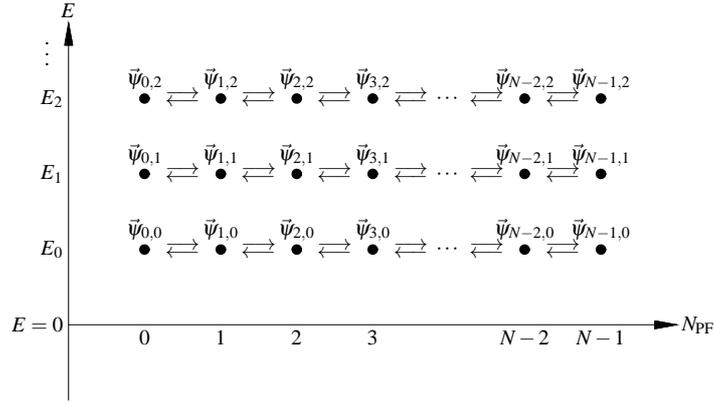} \label{fig:4a}} \hspace{1cm}
\subfigure[Case $M(\alpha)\ell < -1$.]{\input{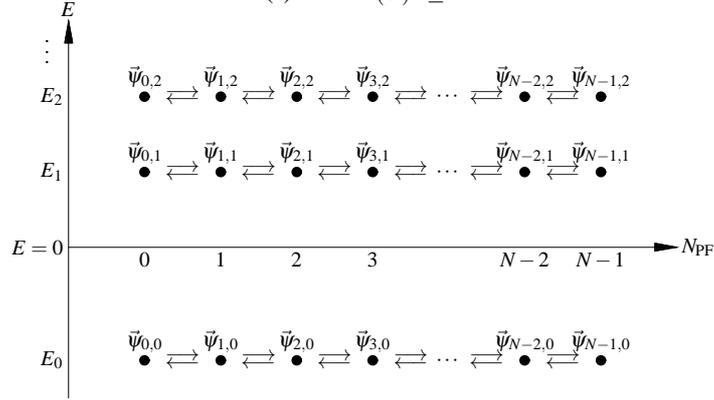} \label{fig:4b}}
\caption{Schematic parasuperspectrum for the type (D, R) and (R, D) boundary conditions. Parasupersymmetry is spontaneously broken.}
\label{fig:4}
\end{figure}

\subsection{Type (D, R) and (R, D) boundary conditions} \label{sec:4.3}
Let us finally solve the model under the type (D, R) and (R, D) boundary conditions.
Since these two boundary conditions are related by the $\mathcal{R}$-transformation, both lead to the isospectral system.
In what follows we will concentrate on the type (D, R) case.

Let us first focus on the case $M(\alpha)\ell \geq -1$ (or $\alpha_{\ast}:=-2\mathrm{arccot}(M_{0}\ell) \leq \alpha \leq \pi$).
In this case all the energy eigenvalues are non-negative and $N$-fold degenerate.
The normalized energy eigenfunctions are given by
\begin{subequations}
\begin{align}
\Vec{\psi}_{n,\nu}(x) &= (-1)^{\frac{n}{2}}\sqrt{\frac{2}{\frac{M(\alpha)}{k_{\nu}^{2} + M^{2}(\alpha)} + \ell}}\sin(k_{\nu}x)\Vec{e}_{n} 	& (n=0,2,4,\cdots,N-2), \label{eq:4.9a}\\
\Vec{\psi}_{n,\nu}(x) &= (-1)^{\frac{n-1}{2}}\sqrt{\frac{2}{\frac{M(\alpha)}{k_{\nu}^{2} + M^{2}(\alpha)} + \ell}}\sin(k_{\nu}(x-\ell))\Vec{e}_{n} 	& (n=1,3,5,\cdots,N-1), \label{eq:4.9b}
\end{align}
\end{subequations}
where $k_{\nu}>0$ ($\nu=0,1,2,\cdots$) is the $(\nu+1)$th positive root of the transcendental equation
\begin{align}
M(\alpha) = -k\cot(k\ell). \label{eq:4.10}
\end{align}
The energy eigenvalues are given by $E_{\nu} = k_{\nu}^{2}$ ($\nu=0,1,2,\cdots$).

\begin{figure}[t]
\centering
\input{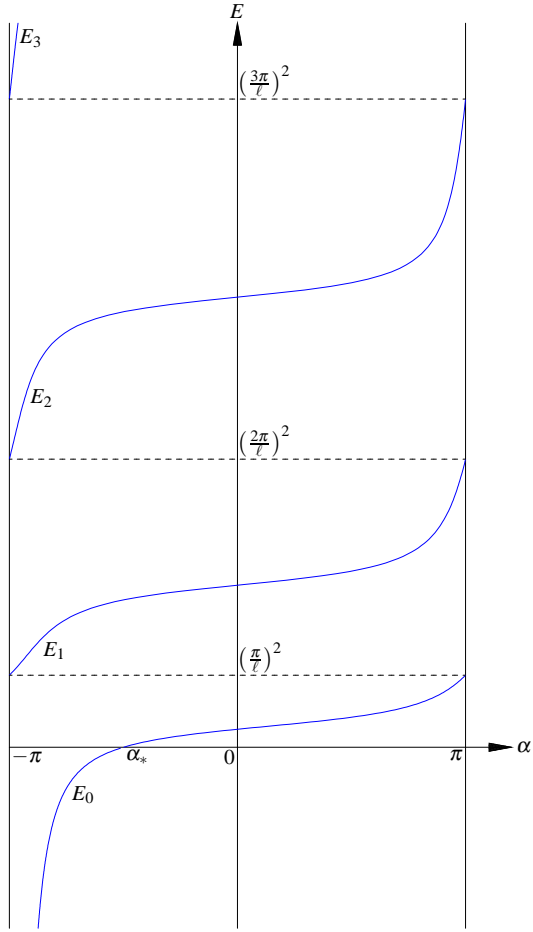}
\caption{Parameter dependence of the energy eigenvalues $\{E_{0}, E_{1}, E_{2}, \cdots\}$ for the type (D, R) and (R, D) boundary conditions. The blue curve $E = E(\alpha)$ is given by the inverse of the phase $\alpha(E) = -2\arctan\bigl(\frac{\sqrt{E}}{M_{0}}\cot(\sqrt{E}\ell)\bigr)$ that follows from Eq. \eqref{eq:4.10}. Spectrum has the period $2\pi$. The ground state energy $E_{0}$ crosses zero at $\alpha = \alpha_{\ast} = -2\mathrm{arccot}(M_{0}\ell)$.}
\label{fig:5}
\end{figure}

Let us next consider the case $M(\alpha)\ell < -1$ (or $-\pi < \alpha < \alpha_{\ast}$).
As shown in Figure \ref{fig:5}, in this case the ground states energy $E_{0}$ becomes negative, while all the excited state energies are remained to be positive.
The ground state wavefunctions are given by replacing $k_{0}$ to $i\kappa$ in \eqref{eq:4.9a} and \eqref{eq:4.9b}
\begin{subequations}
\begin{align}
\Vec{\psi}_{n,0}(x) &= (-1)^{\frac{n}{2}}\sqrt{\frac{2}{\frac{M(\alpha)}{\kappa^{2} - M^{2}(\alpha)} - \ell}}\sinh(\kappa x)\Vec{e}_{n} 	& (n=0,2,4,\cdots,N-2), \label{eq:4.11a}\\
\Vec{\psi}_{n,0}(x) &= (-1)^{\frac{n-1}{2}}\sqrt{\frac{2}{\frac{M(\alpha)}{\kappa^{2} - M^{2}(\alpha)} - \ell}}\sinh(\kappa(x-\ell))\Vec{e}_{n} 	& (n=1,3,5,\cdots,N-1), \label{eq:4.11b}
\end{align}
\end{subequations}
where $\kappa > 0$ satisfies the transcendental equation
\begin{align}
M(\alpha) = -\kappa\coth(\kappa\ell). \label{eq:4.12}
\end{align}
In this case the ground state energy $E_{0}$ is given by $E_{0} = -\kappa^{2} > -M^{2}(\alpha)$.
Opposed to the previous subsections, the ground states are not annihilated by the parasupercharges $Q^{\pm}$.
Hence, in the type (D, R) and (R, D) boundary conditions parasupersymmetry is spontaneously broken.
Figure \ref{fig:4} shows a schematic spectrum valid for both $N$ even and odd.

\section{Conclusions and discussions} \label{sec:5}
It has been long appreciated that supersymmetry and its extensions may be hidden and play a crucial role in quantum mechanics with spectral degeneracy.
In this paper we showed that the Huang-Su parasupersymmetry algebra is hidden behind the degenerate spectra in non-relativistic quantum mechanics for a single free particle on the carambola graph $\mathcal{C}_{N}$.
We imposed the generic $\mathbb{Z}_{N}$ cyclic symmetry on the graph, which plays the role of grading operator $(\mathrm{e}^{2\pi i/N})^{N_{\text{PF}}}$, and obtained the $\mathbb{Z}_{N}$-graded Hilbert space.
We explicitly constructed the parasupercharges and showed that the Huang-Su parasupersymmetry of order $p=N-1$ is hidden behind the $N$-fold degenerate spectrum.
We classified the boundary conditions invariant under parasupersymmetry transformations and found that there are only four types of parasupersymmetry invariant boundary conditions, of which two lead to the parasupersymmetry breaking.
It is interesting to point out here that, in the type (D, R) and (R, D) boundary conditions, in which parasupersymmetry is spontaneously broken, the energy spectrum exhibits spiral structure as shown in Figure \ref{fig:5}.
Consequently, as one varies the parameter $\alpha \in S^{1}$ along $S^{1}$ and completes a cycle,  the energy eigenvalue $E_{\nu}$ gets shifted by unit level, $E_{\nu}(\alpha + 2\pi) = E_{\nu+1}(\alpha)$; in other words, nontrivial Cheon's spiral holonomy \cite{Cheon:1998pt} appears in these cases.
Since all the energy levels are $N$-fold degenerate, the Wilczek-Zee holonomy \cite{Wilczek:1984dh} might also appear under the adiabatic change of parameters that parameterize the basis vectors $\{\Vec{e}_{n}\}$.
It would be interesting to investigate combined phenomena of Cheon's spiral holonomy and the Wilczek-Zee holonomy in this context.

Before closing this paper it should be mentioned about the relation between our results and the previous analysis on hidden supersymmetry structures in quantum mechanics with point interactions \cite{Uchino:2002xb,Nagasawa:2002un,Uchino:2003kh}.
In Ref.\cite{Nagasawa:2002un} Nagasawa \textit{et al.} considered quantum mechanics on $S^{1}$ with two $U(2)$ family of point interactions and studied the hidden $\mathscr{N}=2$ supersymmetry in the spectrum.
Our results include theirs when $N=2$.
In Ref.\cite{Uchino:2002xb} Uchino and Tsutsui studied quantum mechanics on an interval with a single $U(2)$ family of point interactions and revealed the hidden $\mathscr{N} = 1$ and $\mathscr{N} = 2$ supersymmetries in the spectrum.
Our results coincide with their $\mathscr{N}=2$ results when $N=2$ with suitable choice of the basis vectors $\{\Vec{e}_{0},\Vec{e}_{1}\}$.
They further extended to a system on a pair of two intervals each having a single $U(2)$ family of point interactions and showed that large varieties of $\mathscr{N}=2$ and $\mathscr{N} = 4$ supersymmetries are hidden in the spectrum \cite{Uchino:2003kh}.
Basically, our $N=4$ results coincide with their $\mathscr{N}=4$ results when the zero-mode energy eigenvalues are tuned to be degenerate.
The reason why our results on $\mathcal{C}_{N}$ contain the results on these topologically distinct spaces may be understood by topology change \cite{Balachandran:1995jm,Asorey:2004kk,Shapere:2012wn}.
It is also interesting to investigate (para)supersymmetry structures in topologically distinct graphs in a more systematic way.

\bibliographystyle{utphys}
\bibliography{Bibliography}

\end{document}